\documentstyle[11pt,amsmath,amssymb,epsf,epic]{article}

\numberwithin{equation}{section}

\newcommand{\un}{{\mathbb I}}
\newcommand{\ra}{\rightarrow}

\newcommand{\bra}{\langle} 
\newcommand{\ket}{\rangle}
\newcommand{\upket}{|\!\uparrow\rangle}
\newcommand{\downket}{|\!\downarrow\rangle}
\newcommand{\upbra}{\langle\uparrow\!|}
\newcommand{\downbra}{\langle\downarrow\!|}
\renewcommand{\i}{{\rm i}}
\newcommand{\E}{{\mathbb E}}

\newcommand{\be}{\begin{equation}}
\newcommand{\ee}{\end{equation}}
\newcommand{\bea}{\begin{eqnarray}}
\newcommand{\eea}{\end{eqnarray}}
\newcommand{\cx}{{\mathbb C}}
\newcommand{\cz}{{\mathbb Z}}

\newcommand{\ch}{{\cal H}}

\newcommand{\e}{{\rm e}}

\renewcommand{\d}{{\rm d}}

\newcommand{\ffi}{\varphi}

\newcommand{\ep}{\hfill  {\vrule height 10pt width 8pt depth 0pt}}

\newcommand{\grintl}{[\kern-.18em [}
\newcommand{\grintr}{]\kern-.18em ]}

\newcommand{\rx}{{\mathbb R}}

\newcounter{resultcounter}[section]
\renewcommand{\theresultcounter}{\arabic{section}.\arabic{resultcounter}}

\newtheorem{thm}[resultcounter]{Theorem}
\newtheorem{lem}[resultcounter]{Lemma}
\newtheorem{prop}[resultcounter]{Proposition}
\newtheorem{cor}[resultcounter]{Corollary}
\newtheorem{definition}[resultcounter]{Definition}

\def\bed{\begin{definition}}
\def\eed{\end{definition}}

\def\proof{\noindent{\bf Proof.}\ \ }

  \def\cF{{\cal F}}

\newcommand{\R}{{\mathbb R}}
\newcommand{\N}{{\mathbb N}}

\newcommand{\C}{{\mathbb C}}
\newcommand{\Z}{{\mathbb Z}}
\renewcommand{\E}{{\mathbb E}}
\renewcommand{\P}{{\mathbb P}}
\newcommand{\I}{{\mathbb I}}
\newcommand{\T}{{\mathbb T}}
\newcommand{\SS}{{\mathbb S}}

\def\proof{\noindent{\bf Proof.}\ \ }

\newcommand{\fer}[1]{(\ref{#1})}
\newcommand{\scalprod}[2]{\left\langle {#1}, {#2}\right\rangle}
\newcommand{\bbbone}{\mathchoice {\rm 1\mskip-4mu l} {\rm 1\mskip-4mu l}
{\rm 1\mskip-4.5mu l} {\rm 1\mskip-5mu l}}

\begin{document}
\title{Dynamical Localization of Quantum Walks\\
 in Random Environments}
 \author{ Alain Joye\footnote{ Institut Fourier, UMR 5582,
CNRS-Universit\'e de Grenoble I BP 74, 38402 Saint-Martin
d'H\`eres, France.} \footnote{Partially supported by the Agence Nationale de la Recherche, grant ANR-09-BLAN-0098-01}  \and Marco Merkli\footnote{Department of Mathematics, Memorial University of Newfoundland, Canada.}\ \footnote{Partly supported by NSERC Discovery Grant 205247, and by the Institut Fourier through a one-month stay as a professeur invit\'e.}}

\date{ }

\maketitle
\vspace{-1cm}
\abstract{The dynamics of a one dimensional quantum walker on the lattice with two internal degrees of freedom, the coin states, is considered. The discrete time unitary dynamics is determined by the repeated action of a coin operator in $U(2)$ on the internal degrees of freedom followed by a one step shift to the right or left, conditioned  on the state of the coin. For a fixed coin operator, the dynamics is known to be ballistic. 

We prove that when the coin operator depends on the position of the walker and is given by a certain i.i.d. random process, the phenomenon of Anderson localization takes place in its dynamical form. 
When the coin operator depends on the time variable only and is determined by an i.i.d. random process, the averaged motion is known to be diffusive and we compute the diffusion constants for all moments of the position.}


\thispagestyle{empty}
\setcounter{page}{1}
\setcounter{section}{1}


\setcounter{section}{0}

\section{Introduction}

The dynamics of Quantum Walks (QW for short) have become a popular topic in the Quantum Computing community as the simplest quantum generalization of classical random walks, see for example the reviews \cite{AAKV},  \cite{Ke}, \cite{Ko}. In the same way classical random walks play an important role in theoretical computer science, typically in search algorithms, QW provide a natural and fruitful extension in the study of quantum search algorithms, see e.g. \cite{SKW}, \cite{AKR}, \cite{MNRS} and  the review \cite{S} and references therein. 
On the other hand, QW also appeal to physicists interested in quantum dynamics. Indeed, QW can be considered as simple discrete dynamical systems governed by an effective unitary operator, not necessarily given as the exact exponential of $i$ times a microscopic Hamiltonian. For a few models of this type, see e.g. \cite{ADZ}, \cite{M}, \cite{lv},  \cite{bb}, \cite{rhk}, and \cite{ade}, \cite{BHJ}, \cite{dOS}, \cite{HJS} for their mathematical analysis. Moreover, there are recent experimental realizations of QW dynamics:   \cite{Ketal} showed  that cold atoms trapped in optical lattices exhibit a QW for suitably monitored optical lattices and \cite{Zetal} show that the same is true for ions caught in monitored Paul traps.

While several types of QW have been defined and studied in different contexts, we will focus on the simplest one dimensional,  discrete time QW on the lattice, defined in analogy with the classical random walk on the lattice. 
Consider a quantum walker on the lattice carrying two internal degrees of freedom (spin states), called the coin states in this context. The unit time step dynamics is defined by the action of a coin operator in $U(2)$ on the internal degrees of freedom followed by a one step shift to the right or left, conditioned on the state of the coin degree of freedom, see (\ref{a1}) below.
As is well known, when the coin operator is identical at each time step, the QW typically exhibits ballistic dynamics due to translation invariance. Moreover, if the coin operator is chosen at random at each time step, which yields a non autonomous random dynamical system, then, typically, the averaged motion is diffusive, see {\it e.g.} \cite{Ko}, \cite{KBH}, see also \cite{SBBH}.

\medskip

In this paper we address the time-homogeneous case where the coin operator depends on the position of the walker on the lattice and is given by a sequence of random matrices in $U(2)$. This defines a QW in a {\it random environment}.
Such situations were first considered numerically in \cite{KLMW},  \cite{YKE}  for continuous time QW on graphs, that is, for an evolution operators generated by some random discrete Hamiltonian. The outcome of these numerical works provides evidence that Anderson localization, in a strong dynamical form, takes place.  Discrete time QW in random environments were then considered in \cite{Ko1}. By contrast, the author establishes that a certain choice of random coin operators {\it does not} lead to Anderson localization. The random coin operators in \cite{Ko1} are given by a fixed unitary matrix whose diagonal elements carry random phases.

\medskip

We consider here a family of i.i.d. random coin operators characterized by general requirements on the amplitude and transition probabilities to the right and to the left expressed as Assumptions {\bf (a), (b)} and {\bf (c)} below. This family is indexed by a real deterministic parameter and the randomness is determined by i.i.d. phases carried by all matrix elements of the coin operator, see (\ref{defcoin}). We prove that for all values of the deterministic parameter, dynamical localization takes place everywhere in the spectrum, for almost all realizations of coin operators. 

The lack of Anderson localization in the model considered in \cite{Ko1} is explained by the existence of a gauge transformation that fully eliminates the randomness of the model. The set of spatially random unitary matrices we consider coincides with the one considered by  \cite{KBH} in the study of temporal randomness, in one space dimension. 

\smallskip

For completeness, we briefly reconsider the study \cite{KBH} of the non-autonomous case where the coin operators are random in time. We relate the averaged motion to that of a persistent random walk and provide a simple proof of the fact that for all $L\in \N$, the moments of order $2L$, at time $n$, behave as $D(2L)n^L$, for large $n$, with an explicit formula for the diffusion constants $D(2L)>0$.


\section{Setup and Main Results}
\label{sec:setup}
The Hilbert space of pure states 
\begin{equation}
\ch = \cx^2\otimes l^2(\cz)
\label{m1}
\end{equation}
represents two internal degrees of freedom, also called a {\it coin}, with Hilbert space $\cx^2$, and the {\it walker} whose position Hilbert space is $l^2(\cz)$. We fix a canonical basis of $\cx^2$ denoted by $\upket, \downket$, and the position basis consisting of vectors denoted by $|n\rangle$, $n\in\cz$ (eigenvectors of the position operator, the operator of multiplication by the variable $n$). 

The dynamics of the system is composed of discrete steps, each step consisting of a unitary evolution of the coin (operator $C$ on $\cx^2$) followed by the motion of the walker, conditioned on the state of the coin. The latter step is determined by the action
\begin{eqnarray}
\upket\otimes|n\rangle &\mapsto&  \upket\otimes|n+1\rangle \label{m2}\\
\downket\otimes|n\rangle &\mapsto&  \downket\otimes|n-1\rangle \label{m3}
\end{eqnarray}
extended by linearity to $\ch$. This means that if the coin is pointing up the walker will move to the right one step, and if the coin points down the walker moves to the left. The action of \fer{m2}, \fer{m3} is implemented by the unitary operator
\begin{equation*}
S=\sum_{k\in \cz} \left\{ P_\uparrow\otimes|k+1\rangle\langle k| +  P_\downarrow\otimes|k-1\rangle\langle k|\right\}
\end{equation*}
where we have introduced the orthogonal projections
\begin{equation}
P_\uparrow = \upket\upbra \mbox{\quad and \quad} P_\downarrow = \downket\downbra .
\label{m4}
\end{equation}

The one step dynamics consists in tossing the quantum coin and then performing the coin dependent shift
\begin{equation}
\label{a1}
U=S(C\otimes {\mathbb I}) \quad \quad  \mbox{with}\quad \quad 
C=\left[\begin{array}{cc}
a& b\\ c&d
\end{array}\right] \quad \quad \mbox{s.t. } \quad \quad C^*=C^{-1}.
\end{equation}
Hence, if one starts form the state $\upket\otimes|n\rangle$, the (quantum) probability to reach, in one time step, the site $|n+1\rangle$ equals $|a|^2$ whereas that to reach $|n-1\rangle$ equals $1-|a|^2$. Similarly, starting form $\downket\otimes|n\rangle$, the probability to reach the site $|n-1\rangle$ equals $|a|^2$ and that to reach $|n+1\rangle$ is $1-|a|^2$. The evolution operator at time  $n$ reads $U^n$.

Despite the similarity of this dynamics with that of a classical random walk, there is nothing random in the quantum dynamical system at hand. The dynamics is invariant under translations on the lattice $\Z$, which hints at ballistic transport properties. 

\medskip

More precisely, 
let  $X=\un\otimes x$ be the operator defined on its maximal domain in $\cx^2\otimes l^2(\cz)$, where $x$ is the position operator given by $x |k\ket=k|k\ket$, for all $k\in\Z$. For any $L>0$ and any $\Psi$ in the domain of $X^L$, we define
\begin{equation}\label{defxx}
\bra  X^L \ket_{\Psi}(n):=\scalprod{\Psi}{U^{-n} X^L U^{n} \Psi}.
\end{equation}
The analog definition holds for $\bra  |X|^L \ket_{\Psi}(n)$.
We recall in appendix the proof that in our one dimensional setup we have

\begin{lem}[Deterministic walk]
 \label{ballistic} Let $\Psi$ belong to the domain of $X^2$. Then 
$$
\lim_{n\ra\infty}\frac{\bra X^2\ket_{\Psi}(n)}{n^2}=B\geq 0
$$ 
with $B=0$ iff $C$ is off diagonal.
\end{lem}

A QW in a non-trivial environment is characterized by coin operators that depends on the position of the walker: for every $k\in\cz$ we have a unitary $C_k$ on $\cx^2$, and the one step dynamics is given by
\begin{equation}
U = \sum_{k\in \cz} \left\{ P_\uparrow C_k\otimes|k+1\rangle\langle k| + P_\downarrow C_k\otimes|k-1\rangle\langle k|\right\}.
\label{m5}
\end{equation}
We consider a {\it random environment} in which the coin operator $C_k$ is a {\it random} element of $U(2)$, satisfying the following requirements:
\medskip

\noindent
{\bf Assumptions:}

\noindent
{\bf (a)} $\{ C_k\}_{k\in\Z}$ are independent and identically distributed $U(2)$-valued random variables.

\noindent
{\bf (b)} The quantum {\it amplitudes} of the transitions the right and to the left are independent random variables.

\noindent
{\bf (c)} The quantum transition {\it probabilities} between neighbouring sites are deterministic and  independent of the site.

\medskip
The first assumption means that the sites $k\in \cz$ are independent. 
If the walker is localized at site $n$, then the state of the system is of the form $\varphi\otimes|n\rangle$ for some normalized $\varphi\in\cx^2$. The probability amplitudes for transitions to states $\chi\otimes|n+1\rangle$ and $\chi\otimes|n-1\rangle$ ($\chi\in\cx^2$ normalized) are
$\scalprod{\chi}{\uparrow}\scalprod{\uparrow}{C_n\varphi}$ and $\scalprod{\chi}{\downarrow}\scalprod{\downarrow}{C_n\varphi}$, respectively. The second requirement says that these transitions are statistically independent. Finally, the last requirement says that randomness appears as phases (see after \fer{a1}) and that, in a certain sense, we remain close to a classical asymmetric random walk on the lattice.\\

We show in Lemma \ref{invC} below that under assumptions  {\bf (a), (b)} and {\bf (c)}, we can consider without loss of generality 
\begin{equation}\label{defcoin}
C_k=\left[
\begin{array}{cc}
\e^{-\i\omega_k^\uparrow} t & -\e^{-\i\omega_k^\uparrow} r\\
\e^{-\i\omega_k^\downarrow} r & \e^{-\i\omega_k^\downarrow} t
\end{array}
\right], \ \ \ \ \ \mbox{with $0\leq r, t\leq 1 $ and $r^2+t^2=1$}
\end{equation}
and with $\{\omega_k^\uparrow\}_{k\in {\mathbb Z}}\cup\{\omega_k^\downarrow\}_{k\in{\mathbb Z}}$ i.i.d. random variables. 
Thus, the action of the {\it random} operator $U_\omega$ is
\begin{eqnarray}\label{action}
U_\omega\upket\otimes|n\rangle &=&  \e^{-\i\omega^\uparrow_n}t\upket\otimes|n+1\rangle + \e^{-\i\omega^\downarrow_n} r\downket\otimes|n-1\rangle\\ \nonumber
U_\omega\downket\otimes|n\rangle &=&  -\e^{-\i\omega^\uparrow_n}r\upket\otimes|n+1\rangle + \e^{-\i\omega^\downarrow_n} t\downket\otimes|n-1\rangle.
\end{eqnarray}
The elimination of the randomness in the model \cite{Ko1} is a consequence of Lemma \ref{invC} also.

\medskip

To define the random phases properly, we introduce the probability space $(\Omega,{\cal F},\P)$ where $\Omega=\T^{\Z}$, $(\T = \R/2\pi\Z)$, ${\cal F}$ is the $\sigma$-algebra generated by cylinders of Borel sets, and $\P = \otimes_{k\in\Z} \mu$, where $\mu$ is a  probability measure on $\T$. We note expectation values with respect to $\P$ by $\E$. 

The random variables $\omega^\uparrow_k$ and $\omega^\downarrow_k$ on $(\Omega,{\cal F},\P)$ are defined by 
\begin{equation}\label{raph}
\omega^\uparrow_k : \Omega\rightarrow \T, \ \ \ \omega^\uparrow_k =\omega_{2k}\ , \ \ \ \
\omega^\downarrow_k : \Omega\rightarrow \T, \ \ \ \omega^\downarrow_k =\omega_{2k+1}.
\end{equation}
A realization is thus denoted by $\omega=(\cdots, \omega^\uparrow_{-1}, \omega^\downarrow_{-1}, \omega^\uparrow_{0}, \omega^\downarrow_{0}, \omega^\uparrow_{1}, \omega^\downarrow_{1}, \cdots )\in \Omega.$
\medskip

{\bf Main result.\ } Our main result is Theorem \ref{main} and its corollary. Let $U_\omega$ be the one step dynamics of a QW in a random environment defined by (\ref{m5}) with $C_k$, $k\in\Z$ given by (\ref{defcoin}), where 
$\{\omega_k^\#\}_{k\in \N, \#\in\{\uparrow, \downarrow\} }$ are the i.i.d. random variables  defined in (\ref{raph}), distributed according to a measure $\mu$ on $\T$.
\begin{thm}[Spatial disorder]
\label{main}
Suppose that $\mu$ is an absolutely continuous measure $d\mu(\omega)=\tau(\omega)d\omega$, with density $\tau\in L^\infty(\T)$, and that the support of $\mu$ contains a non-empty open set. Then, for any $r\in (0,1)$, there exist $C<\infty$, $\alpha>0$ such that for any $j, k \in \Z$ and any $\sigma, \tau \in \{\uparrow, \downarrow\}$
\begin{equation}\label{loces}
\E \left[ {\sup_{f\in C(\SS),\, \|f\|_\infty\leq 1}}\  \Big|\scalprod{ \sigma \otimes j }{f(U_\omega) \, \tau \otimes k }\Big| \right]\leq Ce^{-\alpha |j-k|}.
\end{equation}
\end{thm} 
{\bf Remarks:}\\
1) The extreme cases $r=1$, $t=0$ and $r=0$, $t=1$ lead to deterministic results that are addressed in the remarks following Lemma \ref{invC}. As seen from (\ref{action}), when $r=0$, the up and down components are independent and propagate respectively to the right and to the left. We get ballistic behaviour for any deterministic choice of phases. When $t=0$, the walker oscillates back and forth, switching its coin state. Hence localization takes place for any deterministic choice of phases.\\
2) Specializing the result to the function $f(z)=z^n$, $z\in \SS$, (\ref{loces}) implies the following almost sure result on the evolution in time of the quantum moments of the QW, see \cite{HJS}.
\begin{cor} There exists a set $\Omega_0$ of probability one, such that for any $\omega\in \Omega_0$, any $L>0$ and for any $\Psi\in \cx^2\otimes l^2(\cz)$ of finite support, there exists $C_{\omega}<\infty$ with
\begin{equation}
\sup_{n\in\Z}\bra |X|^L \ket_{\Psi}(n)<C_{\omega}.
\end{equation}
\end{cor}

\noindent
{\bf Remarks:}\\
1) Another Corollary of Theorem \ref{main} is that the spectrum of $U_\omega$ is pure point, almost surely, see \cite{HJS}.\\
2) The proof of this localization result leans on the fact that the unitary operator $U_\omega$ can be viewed as a doubly infinite five-diagonal band matrix on $l^2(\Z)$, a set of operators first considered in \cite{BHJ}. Moreover, the randomness appears in such a way that it is possible to adapt the Aizenman-Molchanov method \cite{AM} developed for the study of Anderson localization in the self-adjoint case to this unitary setup, along the lines of \cite{HJS}. The random operator at hand differs from those considered in \cite{BHJ, HJS} which forces us to revisit the arguments based on the analysis of products of random transfer matrices and the associated Lyapunov exponents coupled with fractional moment estimates of finite volume restrictions of the associated resolvent.

\medskip

For the sake of comparison, we compute the asymptotics of the expectation of all (integer) moments of the position operator for a dynamics that is random {\it in time}. We follow  \cite{KBH} and consider the random evolution operator at time $n$  given by
\begin{equation}\label{tr}
U_{\omega}(n,0)=U_nU_{n-1}\cdots U_2U_1, \ \ \ \mbox{where} \ \ \ U_k=S(C_k\otimes\I),
\end{equation}
where  the i.i.d. random variables $\{C_k\}_{k\in \Z}$ are defined by (\ref{defcoin}). The random phases are assumed in this case to satisfy 
\begin{equation}\label{exph}
\E(e^{-i\omega_k^\uparrow})=\E(e^{-i\omega_k^\downarrow})=0,
\end{equation}
but their common probability measure $\mu$ is otherwise arbitrary.
We extend the results of  \cite{KBH} on the diffusive behavior of the QW in this setup by proving the following
\begin{prop}[Temporal disorder]
\label{diff}
Assume the evolution operator is given by (\ref{tr}) with random phases  (\ref{raph})
distributed according to a measure $d\mu$ such that  (\ref{exph}) holds. Let $\Psi_0=\ffi_0\otimes |0\ket$ be of norm one. Then, for any $L\in\N$, we have
\bea
\lim_{n\ra\infty}\frac{\E(\bra X^{L}\ket_{\psi_0}(n))}{n^{L/2}}=
D(L)
\eea 
where
\be
D(L)=\left\{ \begin{array}{ll}0 & \mbox{if $L$ is odd}\\
1\cdot 3\cdot 5 \cdots (L-1)(t^2/r^2)^{L/2} & \mbox{if $L$ is even.}\end{array} \right.
\ee
\end{prop}
\medskip

\noindent
{\bf Remark:}\\
As observed in \cite{KBH},  assumption (\ref{exph}) allows one to express  the averaged motion as a {\it persistent} classical random walk on $\Z$, which is known to display a diffusive behaviour. Considering the associated generating function and analyzing its large times asymptotics, we are able to make explicit all diffusion constants. 

\medskip \medskip
Before we turn to the proofs of these results, we briefly investigate the invariance properties of the model stemming from its structure (\ref{m5}) and assumptions {\bf (a), (b)} and {\bf (c)}.

\subsection{Structure of $U_\omega$}

\begin{lem}\label{invC}
Under the assumptions {\bf (a), (b)} and {\bf (c)}, the operator $U_\omega$ defined by (\ref{m5}) is unitarily equivalent to the one defined by the choice
\begin{equation}
 \left[
\begin{array}{cc}
\e^{-\i\omega_k^\uparrow} t & -\e^{-\i\omega_k^\uparrow} r\\
\e^{-\i\omega_k^\downarrow} r & \e^{-\i\omega_k^\downarrow} t
\end{array}
\right] \ \ \ \ \ \ \mbox{\quad where $0\leq t,r\leq 1$ and $r^2+t^2=1$}
\end{equation}
and  $\{\omega_k^\uparrow\}_{k\in {\mathbb Z}}\cup\{\omega_k^\downarrow\}_{k\in{\mathbb Z}}$ are iid random variables defined by (\ref{raph}), 
up to multiplication by a global deterministic phase.
\end{lem}

\proof
We first note that assumption {\bf (a)} implies that we can consider one $C_k$ at a time and {\bf (b)} and {\bf (c)} imply  that the randomness appears as phases only (see after \fer{a1}).  The independence of the right and left probability amplitudes implies that the rows of $C_k$ are independent random variables.

As  $C_k$ must be unitary for all realizations, the scalar product of its columns  equals zero. This shows that the random phases of the elements  on the same line are identical, i.e. 
\begin{equation}
C_k = \left[
\begin{array}{cc}
\e^{-\i\omega_k^\uparrow} a & \e^{-\i\omega_k^\uparrow} b\\
\e^{-\i\omega_k^\downarrow} c & \e^{-\i\omega_k^\downarrow} d
\end{array}
\right]
=
\left[
\begin{array}{cc}
\e^{-\i\omega_k^\uparrow} & 0\\
 0 & \e^{-\i\omega_k^\downarrow} 
\end{array}
\right]
\left[
\begin{array}{cc}
a & b\\
c & d
\end{array}
\right],
\label{m6}
\end{equation}
with
$C\equiv 
\left[
\begin{array}{cc}
a & b\\
c & d
\end{array}
\right]
$
unitary and independent of $k$. 
Parametrizing $C$ as
\begin{equation}
C=e^{-i\theta}\left[
\begin{array}{cc}
te^{-\i\alpha} & \i re^{\i\gamma}\\
\i re^{-\i\gamma} & te^{\i\alpha}
\end{array}
\right], \ \ \ \ \mbox{with $0\leq r,t \leq1$ and $\alpha, \gamma, \theta\in \T$},
\end{equation}
one sees that at the cost of multiplying  $U_\omega$ by $e^{\i\theta}$, we can assume $\theta=0$. Moreover, with 
$\Sigma=\left[
\begin{array}{cc}
1 & 0\\
0 & -\i e^{\i\gamma}
\end{array}
\right]$,
we compute
\begin{equation}
\label{a1''}
\Sigma C\Sigma^{-1}=\left[
\begin{array}{cc}
te^{-\i\alpha} & -r\\
 r & te^{\i\alpha}
\end{array}
\right].
\end{equation}
Up to unitary equivalence, we can assume that $C$ has the form of the r.h.s. of (\ref{a1''}). 
\medskip

To get rid of the last phase $\alpha$, we make use of the representation of $U_\omega$ in the ordered basis $\{\ldots,\upket\otimes|n-1\rangle, \downket\otimes|n-1\rangle, \upket\otimes|n\rangle, \downket\otimes|n\rangle,\ldots\}$ as the band matrix
\begin{equation}
U_\omega= D_\omega S,\mbox{\quad with\quad} S=
\left[
\begin{array}{cccccccc}
\ddots & r & te^{\i\alpha} &   &   & &  &  \\
 & 0 & 0 &  &  &   &  &\\
 & 0 & 0 & r & te^{\i\alpha}  &   &  &\\
 & te^{-\i\alpha}  & -r & 0 & 0 &   &  &\\
 &   &   & 0 & 0 & r & te^{\i\alpha} &\\
 &   &   & te^{-\i\alpha}  & -r & 0 & 0&\\\vspace*{-2.5mm}
 &   &   &   &   & 0 & 0& {}_{\ddots}\\
 &   &   &   &   & te^{-\i\alpha}  & -r & 
\end{array}
\right],
\label{71}
\end{equation}
where (upon relabeling the indices of the random phases) $D_\omega$ is diagonal with i.i.d. entries, $D_\omega={\rm diag}(\ldots,\e^{-\i\omega_k}, \e^{-\i\omega_{k+1}},\ldots)$, and the diagonal of $S$ consists of zeroes. Our convention is to fix a labeling of the canonical basis $e_n$ of that space so that the odd rows contain $r, te^{\i\alpha}$ and the even rows contain $ te^{-\i\alpha}, -r$. 
\medskip

Following \cite{BHJ}, we introduce the unitary operator $V$ defined by $Ve_n=e^{\i\zeta_n}e_n$, where $\zeta_n\in\T$, $n\in \Z$. We compute, for any $k\in\Z$,
\begin{eqnarray}\label{uniteq}
&&V^{-1}U_\omega V e_{2k}=e^{\i(\zeta_{2k}-\zeta_{2k-1})}e^{-\i\omega_{2k-1}}re_{2k-1}+e^{\i(\zeta_{2k}-\zeta_{2k+2})}e^{-\i\omega_{2k+2}}te^{-\i\alpha}e_{2k+2}
\\
&&V^{-1}U_\omega V e_{2k+1}=e^{\i(\zeta_{2k+1}-\zeta_{2k-1})}e^{-\i\omega_{2k-1}}te^{\i\alpha}e_{2k-1}-e^{\i(\zeta_{2k+1}-\zeta_{2k+2})}e^{-\i\omega_{2k+2}}re_{2k+2}.\nonumber
\end{eqnarray}
Choosing for any $k\in\Z$
\begin{equation}
\zeta_{2k+2}=\zeta_{2k+1}=-k\alpha,
\end{equation}
yields the result. \ep

\medskip

\noindent
{\bf Remarks:}\\
1) The last argument of the proof above can be adapted to show that the randomness of  the model of QW in random environment considered by \cite{Ko1} can be gauged away, which explains the absence of localization. Indeed, this model is characterized by random coin matrices of the form (with $t=r=1/\sqrt 2$)
\begin{equation}
\tilde C_k=\left[
\begin{array}{cc}
te^{\i\omega_k} & r\\
r & -te^{-\i\omega_k}
\end{array}
\right], \  \ \mbox{where $\{\omega_k\}_{k\in\Z}$ are i.i.d. random variables on $\T$.
}
\end{equation}
The corresponding operator  $\tilde U_\omega$ then satisfies
\begin{eqnarray}
&&V^{-1}\tilde U_\omega V e_{2k}=e^{\i(\zeta_{2k}-\zeta_{2k-1})}re_{2k-1}+e^{\i(\zeta_{2k}-\zeta_{2k+2})}e^{-\i\omega_{2k}}te_{2k+2}
\nonumber\\
&&V^{-1}\tilde U_\omega V e_{2k+1}=-e^{\i(\zeta_{2k+1}-\zeta_{2k-1})}e^{\i\omega_{2k}}te_{2k-1}+e^{\i(\zeta_{2k+1}-\zeta_{2k+2})}re_{2k+2}.
\end{eqnarray}
Choosing $\zeta_0=0$, $\zeta_{-1}=0$ and, for any $k\in\Z^+$,
\begin{eqnarray}
\zeta_{2k+2}&=&-(\omega_{2k}+\omega_{2k-2}+\cdots+\omega_0),\nonumber\\
\zeta_{-2k}&=&\omega_{-2k}+\omega_{-2k-2}+\cdots+\omega_{-2},\nonumber\\
\zeta_{2k+1}&=&-(\omega_{2k}+\omega_{2k-2}+\cdots+\omega_0),\nonumber\\
\zeta_{-(2k+1)}&=&\omega_{-2k}+\omega_{-2k-2}+\cdots+\omega_{-2}, 
\end{eqnarray}
we see that $\tilde U_\omega $ is unitarily equivalent to the deterministic operator $\tilde U_0 $, characterized by the the deterministic coin operator $\tilde C=\left[
\begin{array}{cc}
t & r\\
r & -t
\end{array}
\right].$
\\
2) We will consider from now on $U_\omega$ to be defined by \fer{71} (with $\alpha=0$) acting on the Hilbert space $l^2(\cz)$, with canonical basis $e_{2k}=|\!\uparrow\rangle\otimes|k\rangle$, $e_{2k+1}=|\!\downarrow\rangle\otimes|k\rangle$. We note that the position operator $x$ on $l^2(\cz)$ is related to $X$ defined by (\ref{defxx}) on ${\mathbb C}^2 \otimes l^2(\cz)$ by $(x-1)/2\leq X\leq x/2$, so that dynamical localization is equivalent in both representations.\\
3) In case $r=0, t=1$, (\ref{71}) shows that $U_\omega$ is equivalent to a direct sum of two shifts, which, in other words, corresponds to $C_k=\un$, $k\in\Z$. This leads to ballistic behaviour. In case $r=1, t=0$, the two-dimensional subspaces span$\{e_{2k-1}, e_{2k}\}$, $k\in\Z$, are invariant, which forbids any kind of transport. 
\\
4) The appealing five diagonal representation \fer{71} of QW on $\N$ is also related so called CMV matrices associated with certain orthogonal polynomials on the unit circle, when restricted to $l^2(\cz^+)$. This fact is used in \cite{CGMV} to study certain properties of deterministic QW.

\bigskip

Finally note that the structure of $S$ can be described also via the two unitary operators (``even'' and ``odd'')
\begin{equation}
B_e = \left[
\begin{array}{cc}
r & t \\
t & -r 
\end{array}
\right]
\mbox{\quad and\quad}
B_o = \left[
\begin{array}{cc}
0 & 1 \\
1 & 0 
\end{array}
\right].
\label{boe}
\end{equation}
$S$ is the product of $S_oS_e$, where $S_e$ is block-diagonal, with blocks $B_e$ placed so that the $(1,1)$ entry of $B_e$ lies on even indices of the diagonal, and $S_o$ is block-diagonal with blocks $B_o$ placed so that the $(1,1)$ entry of $B_o$ lies on odd diagonal elements. Obviously $S_e$ and $S_o$ are unitary, and thus so is $S$ and hence $U_\omega$.

\section{Proofs}

We make use of the structure just described to define finite volume approximations of $S$ and thus $U_\omega$.

\subsubsection{Finite and semifinite volume truncation}
On $l^2(\{2n_0,2m_0\})$ we define the matrices
\begin{equation}
\bar S_o= \left[
\begin{array}{cccc}
1 &     &        &  \\
  & B_o &        &  \\
  &     & \ddots &  \\
  &     &        &  B_o \\
\end{array}
\right],
\qquad
\bar S_e= \left[
\begin{array}{cccc}
B_e &     &        &  \\
  & \ddots &        &  \\
  &     & B_e &  \\
  &     &        &  1 \\
\end{array}
\right],
\end{equation}
where each $1$ is a $1\times 1$ block, and the $2\times 2$ blocks $B_o, B_e$ are given in \fer{boe}. The finite volume propagator is 
\begin{equation}
\bar U_\omega= \bar D_\omega \bar S,\mbox{\quad with\quad} \bar S=\bar S_o\bar S_e=
\left[
\begin{array}{ccccccccccc}
r & t  &   &   &   &   &  & & &  \\
0 & 0  & r  &  t &   &   &  & & &\\
t & -r & 0 & 0 &  &   &  & & & \\
 &   &  0 & 0 & r & t  &  & & & \\
 &   &  t & -r &0  & 0 &  & & & \\
 &   &   &  & 0 & 0 &  & & & \\
 &   &   &  & t & -r  &  & & & \\
 &   &   &   &   &  &\ddots & & & \\
 &   &   &   &   &  &  &r & t & 0\\ 
 &   &   &   &   &  &  &0 & 0 & 0\\ 
 &   &   &   &   &  &  &0 & 0 & 1\\ 
 &   &   &   &   &  &  &t & -r & 0\\ 
\end{array}
\right].
\label{m10}
\end{equation}
Here, $\bar D_\omega = {\rm diag}(\e^{-\i\omega_{2n_0}},\ldots,e^{-\i\omega_{2m_0}})$.

Similarly we define $U^\pm_\omega = D^\pm_\omega S^\pm$ acting on the Hilbert spaces $l^2(\{2n_0,\ldots\})$ and $l^2(\{\ldots,2m_0\})$. The matrix $S^+$ is defined as in \fer{m10}, but where the $2\times4$ blocks are repeated indefinitely towards the bottom right. Similarly we define $S^-$, and the $D_\omega^\pm$ are simply the corresponding semifinite truncations of the diagonal matrix $D_\omega$. Analogous definitions hold for the Hilbert spaces $l^2(\{2n_0+1,\ldots\})$, $l^2(\{\ldots, 2m_0+1\})$.

\subsubsection{Transfer matrix}

{\it Infinite volume.\ }
For $z\in\cx\backslash\{0\}$, consider the equation $(U-z)\psi=0$ (dropping the subscript $\omega$ in the notation). In what follows we always consider $z\neq 0$; an analysis for $z=0$ can be done in a similar way, but is not of any use in this work. Due to the band structure of $U$, we obtain a recurrence relation for the components of vectors solving the above equation. Denote by $\psi_n$ the component of $\psi$ along the basis element $e_n$. A vector $\psi$ solves $(U-z)\psi=0$ if and only if its components satisfy
\begin{equation}
\left[
\begin{array}{c}
\psi_{2n+1}\\
\psi_{2n}
\end{array}
\right] 
= T_z(\omega_{2n}, \omega_{2n-1}) 
\left[
\begin{array}{c}
\psi_{2n-1}\\
\psi_{2n-2}
\end{array}
\right],
\label{m8}
\end{equation}
where the transfer matrix is given by
\begin{equation}
T_z(\omega_{2n}, \omega_{2n-1}) =
\frac{\e^{-\i\omega_{2n}}}{zt}\left[
\begin{array}{cc}
z^2\e^{\i(\omega_{2n-1}+\omega_{2n})}+r^2 & -rt \\
-rt & t^2
\end{array}
\right].
\label{m9}
\end{equation}
We have 
\begin{equation}
\det T_z =\e^{-\i(\omega_{2n}-\omega_{2n-1})},
\label{m67}
\end{equation} 
so equation \fer{m8} can be inverted. It follows that if we choose $\psi_{1}$ and $\psi_{0}$ then all components of $\psi$ are fixed, by \fer{m8}. A priori therefore, each eigenvalue could be doubly degenerate; but this does not happen as we see from the following argument. Suppose that $\psi$ and $\chi$ are two eigenvectors with the same eigenvalue $z$. Then we have 
\begin{equation}
\left[
\begin{array}{cc}
\psi_{2n+1} & \chi_{2n+1}\\
\psi_{2n} & \chi_{2n} 
\end{array}
\right]
= T_z(\omega,n)
\left[
\begin{array}{cc}
\psi_{1} & \chi_{1}\\
\psi_{0} & \chi_{0} 
\end{array}
\right],
\label{m15}
\end{equation}
where
\begin{equation}
T_z(\omega,n) := T_z(\omega_{2n},\omega_{2n-1})\cdots  T_z(\omega_{2},\omega_{1}).
\label{m14}
\end{equation}
The determinant of the matrix on the left side of \fer{m15} tends to zero as $n\rightarrow\infty$, since $\psi$ and $\chi$ are eigenvectors and hence belong to $l^2$. On the other hand, $|\det T_z(\omega,n)|=1$ for all $n$. We conclude that the columns of the matrix to the right in \fer{m15} are multiples of each other, and hence so are the vectors $\psi$ and $\chi$. 

\medskip
\noindent
{\it Solutions $\varphi^\pm$.\ } Take $0\neq z\in\cx$ from the resolvent set of $U$. Then there are unique (up to multiplication by a scalar) solutions $\varphi^\pm \in l^2_\pm$ to $(U-z)\psi=0$. Here, $l^2_\pm = \{\psi\in l^2\ :\ \sum_{n\geq 1}|\psi_{\pm n}|^2<\infty\}$. To see existence we set $\psi_0=(U-z)^{-1}e_0$, so that $(U-z)\psi_0=e_0$, and we choose $\varphi^+_k=[\psi_0]_k$ for $k\geq 1$, $\varphi^-_k=[\psi_0]_k$ for $k\leq -1$
. We then define $\varphi^\pm_k$ for the remaining components $k$ by applying the transfer matrix. Uniqueness of $\varphi^\pm$ is shown just as in the above argument following \fer{m15}.

Similarly to \fer{m15}, we have 
\begin{equation}
\left[
\begin{array}{cc}
\varphi^+_{2n} & \varphi^-_{2n}\\
\varphi^+_{2n-1} & \varphi^-_{2n-1} 
\end{array}
\right]
= T(\omega,z,n)
\left[
\begin{array}{cc}
\varphi^+_2 & \varphi^-_2\\
\varphi^+_1 & \varphi^-_1 
\end{array}
\right],
\label{m24}
\end{equation}
for a matrix satisfying $|\det T(\omega,z,n)|=1$. The columns of the matrix to the right are linearly independent, for otherwise $\varphi^+$ and $\varphi^-$ would be multiples of each other. This in turn would mean that $\varphi^\pm\in l^2$, which is not the case since $z$ is not an eigenvalue of $U$. By taking the determinant in \fer{m24} we see that for all $n\in\mathbb Z$,
\begin{equation}
\varphi_{2n}^+\varphi_{2n-1}^- - \varphi_{2n-1}^+ \varphi_{2n}^-\neq 0.
\label{m25}
\end{equation}

\medskip
\noindent
{\it Semifinite volume.\ } Consider $U^+$ acting on $l^2(\{2n_0,\ldots\})$. We solve the equation $(U^+-z)\psi=0$ recursively and obtain for $n\geq n_0+2$
\begin{equation}
\left[
\begin{array}{c}
\psi_{2n+1}\\
\psi_{2n}
\end{array}
\right] 
= T^+_z(\omega, n)  
\left[
\begin{array}{cc}
-\frac{r}{t} & \quad \frac{z}{t^2}\e^{\i\omega_{2n_0+1}}(z\e^{\i\omega_{2n_0}}-r)\\
1 & 0
\end{array}
\right]
\left[
\begin{array}{c}
\psi_{2n_0+2}\\
\psi_{2n_0}
\end{array}
\right],
\label{m20}
\end{equation}
and  
\begin{equation}\label{ci}
\psi_{2n_0+1}= \frac{\e^{\i\omega_{2n_0}}}{t}(z-r\e^{-\i\omega_{2n_0}})\psi_{2n_0}.
\end{equation}
Here, we have set $T^+_z(\omega,n) = T_z(\omega_{2n},\omega_{2n-1})\cdots  T_z(\omega_{2n_0+4},\omega_{2n_0+3})$. A similar argument as in the infinite volume case shows that each eigenvalue has multiplicity one.

\medskip
\noindent
{\it Lyapunov exponent.\ } Products of transfer matrices yield generalized eigenvectors $\psi$ solutions to $(U-z)\psi=0$ and their asymptotic behaviour at infinity is related to the associated Lyapunov exponents. The following properties of the Lyapunov exponent will, in turn,  be needed in the proof of Lemma \ref{lemma4} below. 

\begin{thm}
\label{thmlyap} Let $\mu$ be absolutely continuous with density $\tau\in L^\infty(\T)$ and having support with nonempty interior. Then,  
there is an $\epsilon>0$ such that for every nonzero $z\in \cx$ with $1-\epsilon< |z| < 1+\epsilon$, the limit
\begin{equation}
\lim_{n\rightarrow\infty} \frac{1}{n} \log \| T_z(\omega_{2n},\omega_{2n-1})\cdots  T_z(\omega_{2},\omega_{1})\| =\gamma(z)
\label{m69}
\end{equation}
exists almost surely, $\gamma(z)$ is deterministic and strictly positive. Moreover, $z\mapsto\gamma(z)$ is continuous.
\end{thm}
{\bf Remark: }The limit $\gamma(z)$ is called the {\it Lyapunov exponent}. \\
The proof of this  theorem is given in appendix.

\subsubsection{Green's function}

{\it Infinite volume.\ } Let $z\neq 0$ be in the resolvent set of $U$, and let $\varphi^\pm$ be the unique solutions of $(U-z)\varphi^\pm=0$ in $l^2_\pm$. We define the resolvent by 
\begin{equation}
G_z^\omega = (U_\omega-z)^{-1}
\label{m26}
\end{equation}
(leaving sometimes out the superscript $\omega$).  
Using the explicit form of $U$ we solve the equation $[(U -z) G_z e_n]_k = \delta_{k,n}$ (Kronecker symbol) for $G_z e_n$ and hence obtain the matrix elements of the resolvent. The following is the result. For all $n\in\mathbb Z$, we have 
\begin{equation}
\scalprod{e_k}{G_z^\omega e_{2n}} = \frac{1}{z} \frac{1}{\varphi_{2n-1}^+\varphi_{2n}^- - \varphi^+_{2n}\varphi_{2n-1}^-}
\left\{
\begin{array}{ll}
\varphi^+_{2n-1}\ \varphi_k^-, & k\leq 2n-1\\
\varphi_k^+\ \varphi_{2n-1}^-, & k\geq 2n
\end{array}
\right.
\label{m27}
\end{equation}
and 
\begin{equation}
\scalprod{e_k}{G_z^\omega e_{2n-1}} = \frac{1}{z}  \frac{1}{\varphi_{2n-1}^+\varphi_{2n}^- - \varphi^+_{2n}\varphi_{2n-1}^-}
\left\{
\begin{array}{ll}
\varphi_{2n}^+\ \varphi_k^-, & k\leq 2n-1\\
\varphi_k^+\ \varphi_{2n}^-, & k\geq 2n
\end{array}
\right.
\label{m28}
\end{equation}
Notice that these matrix elements are well defined due to \fer{m25}.

\bigskip
\noindent
{\it Semifinite volume.\ } One can easily construct a vector $\varphi^a$ with components $\varphi^a_k$, $k\in{\mathbb Z}$, satisfying $(U-z)\varphi^a=0$ and $[(U^+-z)\varphi^a]_k=0$ for all $k\geq a:=2n_0$. Here, $U^+$ is acting on $l^2(\{2n_0,\ldots\})$. The components of $\varphi^a$ are obtained by the transfer matrix for a suitable `initial condition' $\varphi^a_{2n_0}$ and $\varphi^a_{2n_0+1}$. 

In the same way, one can construct $\varphi^b$ with components $\varphi^b_k$, $k\in{\mathbb Z}$, satisfying $(U-z)\varphi^b=0$ and $[(U^--z)\varphi^b]_k=0$ for all $k\leq b:=2m_0$. Here, $U^-$ acts on $l^2(\{\ldots,2m_0\})$. 

The Green's functions on the semi-finite spaces are defined by 
\begin{equation}
G^\pm_z = (U^\pm -z)^{-1},
\label{m33}
\end{equation}
for $|z|\neq 1$ and where the operator is acting on the $l^2$ space over the appropriate half-line of integers. Proceeding as in the infinite-volume case, one shows that $\scalprod{e_k}{G^+_z e_{2n}}$ and   $\scalprod{e_k}{G^+_z e_{2n-1}}$, for $k\geq 2n_0$, are given by the r.h.s. of \fer{m27} and \fer{m28} respectively, where $\varphi^-$ is replaced by $\varphi^a$. Furthermore, $\scalprod{e_k}{G^-_z e_{2n}}$ and   $\scalprod{e_k}{G^-_z e_{2n-1}}$, for $k\leq 2m_0$,  are given by the r.h.s. of \fer{m27} and \fer{m28} respectively, where $\varphi^+$ is replaced by $\varphi^b$.

\bigskip
\noindent
{\it Finite volume.\ } On $l^2(\{2n_0,\ldots, 2m_0\})$ the reduced unitary is given by \fer{m10}. The matrix elements of the Green's function $G_z^{[2n_0,2m_0]} = (U^{[2n_0,2m_0]}-z)^{-1}$ is easily obtained in terms of the vectors $\varphi^a$, $\varphi^b$ defined in the previous paragraph. We mention explicitly only the formula
\begin{equation}
\scalprod{e_{2n_0}}{G^{[2n_0,2m_0]}_z e_{2m_0}} = -\frac{1}{z} \frac{1}{\varphi_{2m_0-1}^a\varphi_{2m_0}^b -\varphi_{2m_0-1}^b \varphi_{2m_0}^a} \varphi^b_{2m_0-1}\varphi^a_{2n_0}.
\label{m33.1}
\end{equation}

\subsection{Bound on fractional moments of Green's function}

For $|z|\neq 1$ denote the matrix elements of the resolvent of $U_\omega$ by
\begin{equation}
\scalprod{e_k}{(U_\omega-z)^{-1}e_l}=G^\omega_z(k,l).
\label{m22}
\end{equation}

\begin{thm}
\label{fmbound}
Assume that the random variables $\omega_k$, $k\in\mathbb Z$, are iid with distribution $\d\mu(\omega_0) = \tau(\omega_0)\d\omega_0$ with $\tau\in L^\infty({\mathbb T})$. Let $0<s<1$ and $0<\epsilon<1$. Then there exists a constant $0<C(s,\epsilon)<\infty$ such that 
\begin{equation}
\E\big[|G^\omega_z(k,l)|^s\big]\leq C(s,\epsilon),
\label{m21}
\end{equation}
for all $z\in \{\zeta\in\cx\ :\ 1-\epsilon <|\zeta|<\infty, |\zeta|\neq 1\}$ and all $k,l\in\mathbb Z$. 
\end{thm}
{\bf Remark:} The result holds for the semifinite volume restrictions $G^\pm_z(k,l)$
as well.\\
The proof of this result is given in Hamza's thesis (Theorem 6.1) and \cite{HJS}. It only depends on the fact that $U_\omega$ is the product $D_\omega S$, where $S$ is unitary and $D_\omega={\rm diag}(\ldots,\omega_k, \omega_{k+1},\ldots)$ has the specified random properties.

\subsection{Exponential decay of fractional moments of Green's function}

\begin{thm}
\label{expfmbound}
Assume that the random variables $\omega_k$, $k\in\mathbb Z$, are iid with distribution $\d\mu(\omega_0) = \tau(\omega_0)\d\omega_0$ with $\tau\in L^\infty({\mathbb T})$ and that the support of $\mu$ contains a non-empty open set. Let $0<\epsilon<1$. There exists an $s$ with $0<s<1/2$, and there exist constants $0<C<\infty$, $\alpha>0$, such that
\begin{equation}
\E\big[|G^\omega_z(k,l)|^s\big]\leq C \e^{-\alpha |k-l|},
\label{m23}
\end{equation}
for all $z\in\cx$ satisfying $|z|\neq 1$ and $\frac{1}{1+\epsilon} <|z| <1+\epsilon$, and for all $k,l\in\mathbb Z$. 
\end{thm}

The proof is done in three steps that we now describe.

\subsubsection{Reduction to even matrix elements}

\begin{thm}
\label{redthm}
Suppose that $|k-l|\geq 4$ and let $m,n$ be the unique integers s.t. $k\in\{2m,2m+1\}$ and $l\in\{2n-1,2n\}$. Then we have 
\begin{equation}
\left| G_z^\omega(k,l)\right| \leq \frac{|z|+t}{r}\left[ \frac{|z|}{r} +\frac{|z|^2+r^2}{rt}\right] \sum_{p_1, \, p_2\in\{0,1\}} |G_z^\omega(2m-2p_1, 2n+2p_2)|.
\label{m29}
\end{equation}
\end{thm}
It follows that for any $s>0$
\begin{equation}
\left| G_z^\omega(k,l)\right|^s \leq C(z,t)^s \sum_{p_1, \, p_2\in\{0,1\}} |G_z^\omega(2m-2p_1, 2n+2p_2)|^s,
\label{m30}
\end{equation}
where the constant $C$ is that in front of the sum in \fer{m29} and where we have we used that $(a+b)^s\leq [a^s +b^s]$ for $a,b\geq 0$ and $0<s<1$.

\medskip
{\it Proof of Theorem \ref{redthm}.\ } We write simply $G$ for $G_z^\omega$ in this proof. We first show that for $k\neq n$,
\begin{equation}
G(2k+1,2n) = \frac{z}{r}\e^{\i\omega_{2k-1}}G(2k-2,2n) -\frac{z^2\e^{-\i(\omega_{2k}+\omega_{2k-1})}+r^2}{rt} G(2k,2n),
\label{m31}
\end{equation}
and that for $k\geq 2n+2$ or $k\leq 2n-1$, 
\begin{equation}
G(k,2n-1) = \frac{z}{r}\e^{+\i\omega_{2n-1}} G(k,2n) -\frac{t}{r}\frac{d_{n+1}}{d_{n}} G(k,2n+2),
\label{m32}
\end{equation}
where $d_n=[\varphi_{2n-1}^+\varphi_{2n}^- -\varphi_{2n}^+\varphi_{2n-1}^-]^{-1}$ (c.f. \fer{m25}). Combining \fer{m31} and \fer{m32}, and using the fact that $|\frac{d_n}{d_{n+1}}|=1$, the bound \fer{m29} is easily obtained. The latter fact follows from 
$$
d_{n+1}^{-1} = -\det 
\left[
\begin{array}{cc}
\varphi^+_{2n+2} & \varphi^-_{2n+2}\\
\varphi^+_{2n+1} & \varphi_{2n+1}^-
\end{array}
\right]
=
 -\det \left(T 
\left[
\begin{array}{cc}
\varphi^+_{2n} & \varphi^-_{2n}\\
\varphi^+_{2n-1} & \varphi_{2n-1}^-
\end{array}
\right]\right)
=d^{-1}_n\det T,
$$
where $|\det T|=1$.

Let us now derive \fer{m32} for $k\leq 2n-1$. We have $G(k,2n-1) = \frac{1}{z}d_n\varphi^+_{2n}\varphi^-_{k}$, and $\varphi^+_{2n}$ is given, using the transfer matrix as in \fer{m8}, by $\varphi_{2n}^+ = \frac{\e^{-\i\omega_{2n}}}{zt}\left( -rt\varphi_{2n-1}^+ +t^2\varphi_{2n-2}^+\right)$. In this relation, we replace $\varphi_{2n-2}^+$ by solving \fer{m8} (using the inverse transfer matrix), yielding the expression $\varphi_{2n-2}^+ = \frac{r}{z}\e^{-\i\omega_{2n-1}}\varphi_{2n+1}^+ +\frac{\e^{-\i\omega_{2n-1}}}{zt}[z^2\e^{\i(\omega_{2n-1}+\omega_{2n})}+r^2]\varphi_{2n}^+$. It follows that 
\begin{eqnarray*}
G(k,2n-1) &=& -\frac{r}{z}\e^{-\i\omega_{2n}}G(k,2n) +\frac{rt}{z^2}\e^{-\i(\omega_{2n}+\omega_{2n-1})} \frac{d_{n+1}}{d_{n}} G(k,2n+2)\\
&& +\frac{1}{z^2}\e^{-\i(\omega_{2n}+\omega_{2n-1})}[z^2\e^{\i(\omega_{2n-1}+\omega_{2n})}+r^2]G(k,2n-1),
\end{eqnarray*}
which yields \fer{m32}. If $k\geq 2n$, the same arguments with $\varphi_{2n}^-$ in place of  $\varphi_{2n}^+$ give the result. Expression \fer{m31} is obtained in an analogous manner.
\hfill $\blacksquare$

\subsubsection{Reduction to finite volume}

\begin{thm}
\label{finvolred}
Assume the hypotheses of Theorem \ref{fmbound}, with $0<s<1/2$. For any pair of integers $m\neq n$ let $k_+=\max\{m,n\}$ and $k_-=\min\{m,n\}$. There exists a constant $C_\mu(s,t, \epsilon) <\infty$ s.t. we have 
\begin{equation}
\E\big[ |G_z(2m,2n)|^s\big]^2 \leq C_\mu(s,t,\epsilon) \E\big[ |G_z^{[2k_-, 2k_+]}(2k_-,2k_+)|^{2s}\big],
\label{m40}
\end{equation}
for all $z$ satisfying $|z|\neq 1$ and $1-\frac{\epsilon}{1+\epsilon} < |z| <1+\epsilon$. 
\end{thm}

{\it Proof.\ } For a given $m$, we decompose $l^2({\mathbb Z}) = l^2(\{\ldots, 2m-1\})\oplus l^2(\{2m,\ldots\})$, and set $U=U^-\oplus U^+ +\Gamma$, where $U^- = D_\omega S_o^{(-\infty,2m-1]} S_e^{(-\infty,2m-1]}$ (of course, the diagonal $D_\omega$ is restricted to the correct half-space) and where $U^+ = D_\omega S_o^{[2m,\infty)} S_e^{[2m,\infty)}$. The operator $\Gamma$ is explicitly given by
\begin{equation}
\scalprod{e_k}{\Gamma e_l}= \Gamma(k,l)=
\left\{
\begin{array}{ll}
r\e^{-\i \omega_{2m-1}},  & (k,l)=(2m-1,2m-1), (2m-1,2m)\\
-r\e^{-\i\omega_{2m}},   & (k,l)=(2m,2m-1), (2m,2m)\\
-t\e^{-\i \omega_{2m-1}}, & (k,l)=(2m-1,2m-2)\\
t\e^{-\i\omega_{2m-1}}, & (k,l)=(2m-1,2m+1)\\
t\e^{-\i\omega_{2m}}, & (k,l)=(2m,2m-2)\\
-t\e^{-\i\omega_{2m}}, & (k,l)=(2m,2m+1)\\
\end{array}
\right.
\label{m41}
\end{equation}
and $\Gamma(k,l)=0$ for all other values $k,l$. Denote 
\begin{equation}
G_z^m:= (U^-\oplus U^+ -z)^{-1} = G_z^{(-\infty,2m-1]}\oplus G_z^{[2m,\infty)},
\label{m43}
\end{equation}
then the second resolvent identity gives $G_z=G_z^m -G_z \Gamma G_z^m$. Suppose that $m\leq n-1$. Then $G_z^m(2m,2n) = G_z^{[2m,\infty)}(2m,2n)$ and we have
\begin{equation}
G_z(2m,2n) = G_z^{[2m,\infty)}(2m,2n) - \sum_{k,l} G_z(2m,k) \Gamma(k,l) G_z^m(l,2n).
\label{m42}
\end{equation}
According to \fer{m41}, the matrix elements $\Gamma(k,l)$ vanish unless $l=2m,2m-1,2m-2,2m+1$. However, for $l=2m-1, 2m-2$ we have $G_z^m(l,2n)=0$ (by the block diagonal form \fer{m43}), so the only terms in the sum in \fer{m42} are with $l=2m,2m+1$. It follows that (still $m<n$)
\begin{eqnarray}
\lefteqn{
|G_z(2m,2n)|}\label{m44}\\
&& \leq \Big\{ 1+2[1+(|z|+r)t^{-1}]\max_{k=2m,2m-1} |G_z(2m,k)|\Big\} |G_z^{[2m,\infty)}(2m,2n)|.
\nonumber
\end{eqnarray}
To arrive at this bound we use in \fer{m42} the relation
$$
G_z^m(2m+1,2n) = \frac{\e^{\i\omega_{2m}}}{t} [z-r\e^{-\i\omega_{2m}}]\  G_z^m(2m,2n)
$$
which follows readily from the explicit expressions of the semi-finite volume Green's function, see (\ref{ci}).

In a next step, we estimate $|G_z^{[2m,\infty)}(2m,2n)|$ in \fer{m44} from above by the finite-volume Green's function $|G_z^{[2m,2n]}(2m,2n)|$. We split the space as $l^2(\{2m,\ldots\infty\}) = l^2(\{2m,2n\})\oplus l^2(\{2n+1,\ldots\infty\})$, so that $U^{[2m,\infty)} = U^{[2m,2n]}\oplus U^{[2n+1,\infty)} +\widetilde\Gamma$, with 
\begin{equation}
\scalprod{e_k}{\widetilde\Gamma e_l}= \widetilde \Gamma(k,l)=
\left\{
\begin{array}{ll}
(r-1)\e^{-\i \omega_{2n-1}},  & (k,l)=(2n-1,2n), \\
t\e^{-\i\omega_{2n-1}},   & (k,l)=(2n-1,2n+1),\\
t\e^{-\i \omega_{2n+2}}, & (k,l)=(2n+2,2n)\\
-(r+1)t\e^{-\i\omega_{2n+2}}, & (k,l)=(2n+2,2n+1)
\end{array}
\right.
\label{m45}
\end{equation}
and $\widetilde\Gamma(k,l)=0$ for all other values $k,l$. By the second resolvent identity, we have $G_z^{[2m,\infty)} = (G_z^{[2m,2n]}\oplus G_z^{[2n+1,\infty)}) (1-\widetilde\Gamma G_z^{[2m,\infty)})$. Taking the matrix elements, with $m<n$, we obtain
\begin{eqnarray*}
\lefteqn{
G_z^{[2m,\infty)}(2m,2n)}\\
&& = G_z^{[2m,2n]}(2m,2n) - \sum_{k,l}\scalprod{e_{2m}}{G_z^{[2m,2n]}\oplus G_z^{[2n+1,\infty)}e_k} \widetilde \Gamma(k,l) G_z^{[2m,\infty)}(l,2n).
\end{eqnarray*}
Note that due to \fer{m45}, the sum over $k$ is really only over $k=2n-1, 2n+2$, and the term with $k=2n+2$ is absent since the scalar product in the sum vanishes for this value of $k$. By using additionally that $G^{[2m,2n]}(2m,2n-1) = z\e^{\i\omega_{2n-1}} G^{[2m,2n]}(2m,2n)$, which follows from the explicit formulas for the Green's function, we obtain the bound
\begin{equation}
|G_z^{[2m,\infty)}(2m,2n)|\leq \left\{ 1+3|z|\max_{k=2n,2n+1} |G_z^{[2m,\infty)}(k,2n)|\right\} |G_z^{[2m,2n]}(2m,2n)|.
\label{m46}
\end{equation}
Combining \fer{m46} with \fer{m44} yields that for $m\leq n-1$, 
\begin{eqnarray}
\lefteqn{
|G_z(2m,2n)| \leq \Big\{ 1+2(1+(|z|+r)t^{-1})\max_{k=2m,2m-1}|G_z(2m,k)| \Big\} }\nonumber\\
&\times& \big\{ 1+3|z|\max_{k=2n,2n+1} |G_z^{[2m,\infty)}(k,2n)|\Big\} |G_z^{[2m,2n]}(2m,2n)|.\qquad
\label{m47}
\end{eqnarray}
We take the expectation of the inequality \fer{m47}, use H\"older's inequality and Theorem \ref{fmbound} to arrive at the bound ($m\leq n-1$)
\begin{equation}
\E\big[|G_z(2m,2n)|^s\big]^2 \leq C_\mu(s,t, \epsilon,z)\  \E\big[ |G_z^{[2m,2n]}(2m,2n)|^{2s}\big],
\label{m48}
\end{equation}
for $0<s<1/2$, $0<\epsilon<1$, $|z|\neq 1$ s.t. $1-\epsilon<|z|<\infty$, and where $C_\mu(s,\epsilon,z)$ is bounded uniformly in compact regions of $z$. 

Next we deal with matrix elements of the resolvent with $m\geq n+1$. One can proceed as in the above argument, or instead use the following path. Since $U$ is unitary, 
$|G_z(k,l)| = |\scalprod{e_k}{(U-z)^{-1}e_l}| = |\scalprod{e_l}{(U^{-1}-\bar z)^{-1} e_k}|$, and $(U^{-1}-\bar z)^{-1} =-1/\bar z -(1/\bar z)^2 (U-1/\bar z)^{-1}$. Hence we have for $m\geq n+1$
\begin{equation}
|G_z(2m,2n)| = \frac{1}{|z|^2} |G_{1/\bar z}(2n,2m)|.
\label{m49}
\end{equation}
We want to use the bound \fer{m48} on the right hand side of \fer{m49}. The condition $1-\epsilon <1/|\bar z| <1+\epsilon$ is equivalent to $1-\frac{\epsilon}{1+\epsilon} <|z|<1+\frac{\epsilon}{1-\epsilon}$. It follows that under the last condition on $|z|$ we have, for $m\geq n+1$,
\begin{equation}
\E\big[|G_z(2m,2n)|^s\big]^2 \leq C_\mu(s,t,\epsilon,1/\bar z)\ \frac{1}{|z|^{4s}}\E\big[ |G_{1/\bar z}^{[2n,2m]}(2n,2m)|^{2s}\big].
\label{m50}
\end{equation}
Combining \fer{m48} and \fer{m50} yields the bound \fer{m40}, provided $|z|\neq 1$, $1-\epsilon<|z|<1+\epsilon$ and $1-\frac{\epsilon}{1+\epsilon} <|z| < 1+\frac{\epsilon}{1-\epsilon}$. \hfill $\blacksquare$

\subsubsection{Exponential decay in finite volume}

We assume that the support of the measure $\mu$ contains a non-empty open set in $[0,2\pi)$. This implies positivity of the Lyapunov exponent, see Theorem \ref{thmlyap}, a result we use implicitly in this section.

\begin{thm}
\label{finvoldecthm}
There are numbers $\alpha, s$ satisfying $\alpha>0$, $0<s<1$ such that for all $z\in\cx$, $|z|\neq 0,1$ and $n<m$, we have 
\begin{equation}
\E\left[ |G_z^{[2n,2m]}(2n,2m)|^s\right] \leq C \e^{-\alpha(m-n)}.
\label{m80}
\end{equation}
The constant $C$ depends on $s,z,t$ and $\mu$, but not on $m,n$. It is furthermore uniform in $z$, for $|z|\neq 1$ restricted to compact sets of $\C\setminus\{ 0\}$   (see explicit bound in proof).
\end{thm}

{\it Proof of Theorem \ref{finvoldecthm}.\ } The proof is based on the following two Lemmas.

\begin{lem}
\label{lemma2}
Let $0<s<1$. Then there is a constant $0<C_\mu(s)<\infty$ s.t. for all $|z|\neq 0,1$, 
\begin{equation}
\E\big[ |G_z^{[2n,2m]}(2n,2m)|^s \big] \leq |z|^{-s} (1+|z|^{-s})(1+2^sC_\mu(s))\  \E\left[
\left\|
\begin{array}{cc}
\varphi_{2m}^a\\
\varphi_{2m-1}^a
\end{array}
\right\|^{-s}
\right].
\label{m51}
\end{equation}
Here, $\varphi^a$ is the solution satisfying the left boundary condition at $a=2n$, see also before \fer{m33}. 
\end{lem}

We present a proof of this Lemma at the end of the present section. The following a priori bound has been adapted from the self-adjoint Anderson model situation (see \cite{CKM}, Lemma 5.1), and is given in Appendix A of \cite{H}. The proof uses strict positivity and continuity of the Lyapunov exponent, which we prove in Theorem \ref{thmlyap} below. 

\begin{lem}
\label{lemma4}
Let $\Lambda\subset \cx$ be a compact set not containing the origin. There are numbers $\alpha>0$ and $0<s<1$ (depending on $\Lambda$), such that the product of transfer matrices \fer{m9} satisfy, for all $m>n$,
\begin{equation}
\E\left[ \|T_z(\omega_{2m},\omega_{2m-1})\cdots T_z(\omega_{2n},\omega_{2n-1}) v\|^{-s} \right] \leq \bar C\e^{-\alpha(m-n)},
\label{m82}
\end{equation}
for any normalized vector $v\in\cx^2$. The constant $\bar C$ is independent of $z\in\Lambda$ and $v$. 
\end{lem}

A combination of these two Lemmas yields a proof of Theorem \ref{finvoldecthm} as follows. Solving the equation \fer{m8} for $\varphi^a_{2m-1}$ gives the relation $\varphi^a_{2m-1} = \e^{-\i\omega_{2m-1}}[t\varphi_{2m+1}^a +r\varphi_{2m}^a]/z$, and therefore
$$
\left[
\begin{array}{c}
\varphi_{2m+1}^a\\
\varphi_{2m}^a
\end{array}
\right]
= \frac1t
\left[
\begin{array}{cc}
-r & z\e^{\i\omega_{2m-1}} \\
t & 0
\end{array}
\right]
\left[
\begin{array}{c}
\varphi_{2m}^a\\
\varphi_{2m-1}^a
\end{array}
\right]\equiv \tilde T \left[
\begin{array}{c}
\varphi_{2m}^a\\
\varphi_{2m-1}^a
\end{array}
\right].
$$
We have $\| \tilde T\|\leq \widetilde C(1+|z|)/t$, for some $\widetilde C$ independent of $z,r,t$ and any of the phases.  
Using this expression and definition of the transfer matrices \fer{m8} we obtain
\begin{equation}
\left\|
\left[
\begin{array}{c}
\varphi_{2m}^a\\
\varphi_{2m-1}^a
\end{array}
\right]
\right\| \geq 
\big\|
T_z(\omega_{2m},\omega_{2m-1})\cdots T_z(\omega_{2n+2},\omega_{2n+1}) v^a
\big\| \frac{t}{\widetilde C (1+|z|)}
\label{m81}
\end{equation}
where we chose $\left[
\begin{array}{c}
\varphi_{2n+1}^a\\
\varphi_{2n}^a
\end{array}
\right]$ to be the normalized vector
$$
v^a = \frac{1}{\sqrt{t^2+|z\e^\i\omega_{2n}-r|^2}} \left[
\begin{array}{c}
z\e^{\i\omega_{2n}}-r
\\
t
\end{array}
\right] .
$$
Combining \fer{m81} with \fer{m51} and \fer{m82} proves the bound \fer{m80} and hence Theorem \ref{finvoldecthm}. It remains to give the 

\medskip
{\it Proof of Lemma \ref{lemma2}.\ } The components of $\varphi^b$, with $b=2m$, satisfy $-z\varphi_{2m-1}^b + \e^{-\i\omega_{2m-1}}\varphi^b_{2m}=0$, see before \fer{m33}. Moreover, since $\varphi^b$, is defined modulo a multiplicative factor only, we set $\varphi_{2m-1}^b=1$ (note that $\varphi_{2m-1}^b=0$ would imply that $\varphi^b=0$, so the normalization $\varphi_{2m-1}^b=1$ is possible). Therefore we obtain from \fer{m33.1} the following expression for Green's function:
\begin{equation}
G_z^{[2n,2m]}(2n,2m) = \frac{1}{z} \frac{1}{\varphi^a_{2m} - z\e^{\i\omega_{2m-1}}\varphi_{2m-1}^a}.
\label{m52}
\end{equation}
Note that $\varphi_{2m}^a$ and $\varphi_{2m-1}^a$ cannot both vanish, since otherwise we would have $\varphi^a=0$. 
It follows from \fer{m8} that the components of $\varphi^a$ satisfy
\begin{equation}
\left[
\begin{array}{c}
\varphi^a_{2m-1}\\
\varphi^a_{2m-2}
\end{array}
\right] 
= T_z(\omega_{2m-2}, \omega_{2m-3}) 
\left[
\begin{array}{c}
\varphi^a_{2m-3}\\
\varphi^a_{2m-4}
\end{array}
\right],
\label{m53}
\end{equation}
and furthermore, that $\varphi_{2m}^a =\frac{\e^{-\i\omega_{2m}}}{z}\left[ t\varphi_{2m-2}^a -r\varphi_{2m-1}^a\right]$. Consequently, $\varphi_{2m}^a$ depends only on $\omega_{2m}$ and $\omega_j$, with $j\leq 2m-2$. We have
\begin{eqnarray}
\lefteqn{
\E\left[ \left|G_z^{[2n,2m]}(2n,2m)\right|^s\right]}\nonumber\\
&&
= |z|^{-s} \widehat\E\left[ \int_0^{2\pi} \d\mu(\omega_{2m-1}) \frac{1}{|\varphi^a_{2m} -z\e^{\i\omega_{2m-1}}\varphi^a_{2m-1}|^s}\right],
\label{m54}
\end{eqnarray}
where $\widehat \E$ is the expectation over all $\omega_j$, $j=2n,\ldots,2m-2$ and $j=2m$. The dependence on $\omega_{2m-1}$ of the integrand in
\begin{equation}
I:= \int_0^{2\pi} \d\mu(\omega_{2m-1}) \frac{1}{|\varphi^a_{2m} -z\e^{\i\omega_{2m-1}}\varphi^a_{2m-1}|^s}
\label{m55}
\end{equation}
is concentrated exclusively in $\e^{\i\omega_{2m-1}}$. Let us define the vector
\begin{equation}
v_m := 
\left[\begin{array}{c}
\varphi_{2m}^a \\
\varphi_{2m-1}^a
\end{array}\right].
\label{m56}
\end{equation}
If $\varphi_{2m-1}^a=0$ then we have $I =|\varphi_{2m}^a|^{-s} = \|v_m\|^{-s}$. 
If $\varphi_{2m}^a=0$ then we have $I =|z|^{-s} |\varphi_{2m-1}^a|^{-s} =  |z|^{-s} \|v_m\|^{-s}$.
Next suppose that both $\varphi_{2m-1}^a$ and $\varphi_{2m}^a$ are nonzero. We distinguish two cases:  either $|\varphi_{2m}^a|\geq |\varphi_{2m-1}^a|$ or $|\varphi_{2m}^a|<|\varphi_{2m-1}^a|$. In the former case we have $\|v_m\| \leq |\varphi_{2m}^a| + |\varphi_{2m-1}^a|\leq 2 |\varphi_{2m}^a|$, 
and
\begin{eqnarray}
I&= &|\varphi^a_{2m}|^{-s}\int_0^{2\pi} \d\mu(\omega_{2m-1}) \left| \e^{-\i\omega_{2m-1}} - z\varphi^a_{2m-1}/\varphi^a_{2m}\right|^{-s}\nonumber\\
&&  \leq  C_\mu(s) |\varphi_{2m}^a|^{-s} \leq 2^sC_\mu(s) \|v_m\|^{-s}.
\label{m57}
\end{eqnarray}
Here, we have used that for all $0<s<1$ there exists $0<C_\mu(s)<\infty$ such that for all $\beta\in\cx$
$$
\int_0^{2\pi} \d\mu(\omega) |\e^{\pm i\omega} -\beta|^{-s} \leq C_\mu(s),
$$
see \cite{J}. Next we consider the case $|\varphi_{2m}^a|<|\varphi_{2m-1}^a|$, in which case we have $\|v_m\|\leq |\varphi_{2m}^a| + |\varphi_{2m-1}^a|\leq 2|\varphi_{2m-1}^a|$, and 
\begin{eqnarray}
I&= &|\varphi^a_{2m-1}|^{-s} |z|^{-s} \int_0^{2\pi} \d\mu(\omega_{2m-1}) \left| \e^{\i\omega_{2m-1}} - z^{-1}\varphi^a_{2m}/\varphi^a_{2m-1}\right|^{-s}\nonumber\\
&&  \leq  |z|^{-s} C_\mu(s) |\varphi_{2m-1}^a|^{-s} \leq |z|^{-s} 2^s C_\mu(s) \|v_m\|^{-s}.
\label{m58}
\end{eqnarray}
Combining these estimates, we see that in any event,
\begin{equation}
I\leq   (1+|z|^{-s})(1+2^sC_\mu(s)) \|v_m\|^{-s}.
\label{m59}
\end{equation}
This bound, together with \fer{m54}, yields \fer{m51} with $\widehat\E$ instead of $\E$. But both expressions are the same since $v_m$ does not depend on $\omega_{2m-1}$.

This completes the proof of Lemma \ref{lemma2} and with that the proof of Theorem \ref{finvoldecthm}. \hfill $\blacksquare$

\subsection{Proof of Theorem \ref{expfmbound}}

Let $s, \alpha, C$ be as in Theorem \ref{finvoldecthm}. Combining the latter theorem with Theorem \ref{finvolred}, we obtain the bound
\begin{equation}
\E\left[ |G_z(2n,2m)|^{s/2}\right] \leq C_1 \e^{-2\alpha|m-n|},
\label{m86}
\end{equation}
for all $|z|\neq 0,1$ satisfying the bound indicated in Theorem \ref{finvolred}. The constant $C_1$ and all further constants $C_j$ introduced in this proof depend on $z,s,\epsilon$ and $t$ and are uniform in $z$ in compacts of $\C\setminus\{ 0\}$. We use the inequality \fer{m30} after Theorem \ref{redthm} to arrive at the following bound for $|k-l|\geq 4$,
\begin{equation}
\E\left[ |G_z(k,l)|^{s/2}\right] \leq C_2\sum_{p_1, p_2\in\{0,1\}}  \E\left[ |G_z(2m-2p_1,2n+2p_2)|^{s/2}\right].
\label{m87}
\end{equation}
Next, since $|m-p_1-(n+p_2)|\geq |m-n|-2$, and $|k-l|\leq |2m-2n|+2$, combining \fer{m86} and \fer{m87} gives 
\begin{equation}
\E\left[ |G_z(k,l)|^{s/2}\right] \leq C_3 \e^{-\alpha|k-l|},
\label{m88}
\end{equation}
provided $|k-l|\geq 4$ and $|z|\neq 1$, $\frac{1}{1+\epsilon}<|z|<1+\epsilon$. Finally, if $|k-l|<4$, the bound \fer{m88} is implied by Theorem \ref{fmbound}. This completes the proof of Theorem \ref{expfmbound}. \hfill $\blacksquare$

\subsection{Proof of Theorem \ref{main}}

Exponential decay of fractional moments of Green's function implies dynamical localization for band matrices of the type $U_\omega=D_\omega S$, as shown in the following result.

\begin{thm}[\cite{HJS}, Theorem 3.2] 
\label{hjsthm}
Assume that the random variables $\omega_k$ satisfy the conditions of Theorem \ref{main} and that for some $s\in(0,1)$, $C<\infty$, $\alpha>0$ and $\epsilon>0$,
$$
\E(|G_z(k,l)|^s)\leq C\e^{-\alpha|k-l|}
$$
for all $k,l\in \mathbb Z$ and all $z\in\cx$ s.t. $1-\epsilon < |z|<1$. Then there is a $\widetilde C<\infty$ such that
$$
\E\left[ \sup_{f\in C({\mathbb S}), \|f\|_\infty\leq 1} \big|\scalprod{e_k}{f(U_\omega)e_l}\big|\right]\leq \widetilde C \e^{-\alpha|k-l|/4}
$$
for all $k,l\in\mathbb Z$. 
\end{thm}
With the identification of the basis elements $e_{2k}=|\uparrow\rangle\otimes|k\rangle$, $e_{2k+1}=|\downarrow\rangle\otimes|k\rangle$, see after \fer{71}, we immediately obtain that Theorems \ref{expfmbound} and \ref{hjsthm} imply Theorem \ref{main}.\hfill $\blacksquare$

\section{Temporal disorder}
For the sake of comparison, we briefly consider in this section the case where the disorder is introduced in the model through the time variable. This means that at each time step, the evolution of the coin variable is randomly chosen within a set of iid unitary $2\times 2$ matrices $\{C_k\}_{k\in \Z}$. Therefore, the random evolution after $n$ steps reads
\be
U(n,0)=U_nU_{n-1}\cdots U_2U_1, \ \ \ \mbox{where} \ \ \ U_k=S(C_k\otimes\I).
\ee

As we will see, our choice of random unitary matrices $\{C_k\}_{k\in \Z}$ 
\begin{equation}
C_k = \left[
\begin{array}{cc}
\e^{-\i\omega_k^+} t & -\e^{-\i\omega_k^+} r\\
\e^{-\i\omega_k^-} r & \e^{-\i\omega_k^-} t
\end{array}
\right]
\label{trc}
\end{equation}
with $\{\omega_k^+\}_{k\in {\mathbb Z}}\cup\{\omega_k^-\}_{k\in{\mathbb Z}}$ iid subjected to the condition
\be
\label{zcor}
\E(\e^{\i \omega_k^+})=\E(\e^{\i \omega_k^-})=0,
\ee 
naturally leads to the study of (classical) {\it persistent random walks}. For notational reasons which will be clear below, we change notations to $\uparrow\ =+1, \downarrow \ =-1$ so that
\begin{equation}
\upket = |+1\ket, \ \ \ \downket =|-1\ket, \ \ \ P_\uparrow=P_{+1}, \ \ \ P_\downarrow=P_{-1}.
\end{equation}

\medskip

We first state a deterministic result dealing with expectation values of position operators at time $n$.\\

Let  $X$ denote the position operator on $\C^2\otimes l^2(\Z)$ defined by $(X\psi)(x)=(\I\otimes x)\psi(x)$, $x\in\Z$, on its maximal domain 
$$
D=\left\{\psi\in \C^2\otimes l^2(\Z), \mbox{\ s.t.}\, \sum_{x\in\Z}\sum_{\sigma\in\{+1,-1\}}\|(P_\sigma\otimes x)\psi(x)\|_{\C^2}^2<\infty\right\}.
$$ 
For $f: \Z\mapsto \C$, we define the operator $F(X)$ on  $\C^2\otimes l^2(\Z)$ by $(F\psi)(x)=(\I\otimes f(x))\psi(x)$ on its  maximal domain $D_F$ via the spectral Theorem. For any $\Psi_0\in \C^2\otimes l^2(\Z)$ such that $U(n,0)\psi_0\in D_F$, we note
\be
\scalprod{\Psi_0}{U(n,0)^*F(X)U(n,0)\psi_0} \equiv\bra F(X)\ket_{\psi_0}(n).
\ee

Explicit computations with $P_{\pm 1}=|\pm 1\ket\bra\pm 1|$ yield the following Lemma:
\begin{lem} With the notations above,
\bea
U_nU_{n-1}\cdots U_1&=&\sum_{\sigma_n,\cdots, \sigma_{1}\atop \sigma_j\in \{-1,1\}}\sum_{x\in\Z}P_{\sigma_n}C_n\cdots P_{\sigma_1}C_1\otimes |x+\sum_{j=1}^n\sigma_j\ket\bra x|\nonumber\\
&=&\sum_{x\in\Z}\sum_{k=-n}^nJ_k(n)\otimes |x+k\ket\bra x|
\eea
where
\be
J_k(n)=\sum_{\sigma_n,\cdots, \sigma_{1}\atop \sum_{j=1}^n\sigma_j=k}P_{\sigma_n}C_n\cdots P_{\sigma_1}C_1 \in M_{2}(\C)
\ee
satisfies $J_k(n)=0$ if $k$ and $n$ have different parities or if $k>n$ or $k<-n$.\\

Moreover, if $\Psi_0=\ffi_0\otimes |0\ket$, we have for any $f: \Z\mapsto \C$ and all $n\in\N$,
\be
\bra F(X)\ket_{\psi_0}(n)=\sum_{k=-n}^nf(k)\scalprod{\ffi_0}{J_k^*(n)J_k(n)\ffi_0}_{\C^2},
\ee
where
$W_k(n)\equiv \scalprod{\ffi_0}{J_k^*(n)J_k(n)\ffi_0}_{\C^2}$ satisfies
\be
W_k(n)\geq 0 \ \ \ \mbox{and} \ \ \ \sum_{k=-n}^n W_k(n)=\|\ffi_0\|^2.
\ee
\end{lem}
{\bf Remarks:}\\
i) If the initial coin vector $\ffi_0$ is normalized, which we assume from now on, then the quantum mechanical expectation value of the operator $F(X)$ at time $n$ coincides with the expectation value of the function $f$ with respect to the classical discrete probability distribution $\{ W_k(n)\}_{k\in \{-n,\ldots, n\}}$ on $\{-n,\ldots ,n\}\subset \Z$. The quantity $W_k(n)$ is interpreted as the probability to reach site $k\in \Z$ in $n$ steps.\\
ii) The probabilities $\{ W_k(n)\}_{k\in \{-n,\ldots, n\}}$ are actually $\ffi_0$-dependent random variables for randomly chosen coin operators $\{C_k\}_{k\in\Z}$ with arbitrary distribution. Taking expectation with respect to the distribution of the $C_k$'s, we get a new discrete probability distribution on $\{-n,\ldots ,n\}\subset \Z$ given by $\{w_k(n)\}_{k\in \{-n,\ldots, n\}}$ with 
\be
w_k(n)=\E(W_k(n)), \ \   \ k\in \{-n,\ldots, n\}, \ \ \forall\ n\in \N,
\ee
with same interpretation in terms of a classical random walk on $\Z$.

\medskip

We shall focus on the expectation value of the quantum mechanical moments of the position operator for our choice \fer{trc} of $\{C_k\}_{k\in\Z}$ under condition \fer{zcor}, for a normalized initial condition of the form $\psi_0=\ffi_0\otimes |0\ket$, i.e. on
\be
\E(\bra X^{L}\ket_{\psi_0}(n))=\sum_{k=-n}^nk^Lw_k(n) \ \ \mbox{as }\ \ n\ra \infty.
\ee

We remind the reader that a {\it persistent} or {\it correlated random walk} on $\mathbb Z$ is determined by a probability $p$ that the walker moves one unit in the same direction as in the previous step, and a probability $1-p$ that the walker moves one unit in the opposite direction to the last step. See e.g. \cite{rh}.

We let $S_n=\sum_{j=1}^n\sigma_j$ with $\sigma_j\in\{-1,+1\}$ be the random walk defined by $\P(S_n=k)=w_k(n)$. This random walk is characterized as follows.

\begin{prop}\label{prw}
Assume the matrices $\{C_k\}_{k\in\Z}$ are given by \fer{trc} and \fer{zcor} holds. Take $\psi_0=\ffi_0\otimes |0\ket$ with $\ffi_0=\alpha|+\ket + \beta|-\ket\in\C^2$ normalized. Then $S_n$ is a {\em persistent random walk} with parameters
\begin{eqnarray*}
&&\P(\sigma_1=+1)=|\alpha|^2t^2+|\beta|^2r^2-2\Re(\overline{\alpha}\beta)rt:=a\\
&&\P(\sigma_1=-1)=|\alpha|^2r^2+|\beta|^2t^2+2\Re(\overline{\alpha}\beta)rt:=b=1-a\\
&&\P(\sigma_j=+1|\sigma_{j-1}=+1)=P(\sigma_j=-1|\sigma_{j-1}=-1)=t^2\\
&&\P(\sigma_j=-1|\sigma_{j-1}=+1)=P(\sigma_j=+1|\sigma_{j-1}=-1)=r^2=1-t^2,
\end{eqnarray*}
for $j\geq 2$.
\end{prop}
{\bf Proof:} For $n\in\N$, let $\sigma=(\sigma_1,\sigma_2,\ldots,\sigma_n)\in\{-1,+1\}^n$. We need to consider 
\be\label{comp}
w_k(n)=\sum_{\sigma, \sigma' \in \{-1,+1\}^n\atop \sum_{j=1}^n\sigma_j=\sum_{j=1}^n\sigma'_j=k }\E(\scalprod{\ffi_0}{C_1^*P_{\sigma'_1}C_2^*P_{\sigma'_2}\cdots C_n^*P_{\sigma'_n}P_{\sigma_n}C_n\cdots
P_{\sigma_1}C_1\ffi_0}),
\ee
where $\sigma_n=\sigma'_n$. With $P_\sigma=|\sigma\ket\bra\sigma|$, $\sigma=\pm 1$, and reorganizing the product, the scalar product under the sum equals 
$$
\overline{\scalprod{\sigma'_1}{C_1\ffi_0}}\scalprod{\sigma_1}{ C_1\ffi_0}\prod_{j=2}^n
\overline{\scalprod{\sigma'_j}{ C_j\sigma'_{j-1}}}\scalprod{\sigma_j}{ C_j\sigma_{j-1}},
$$
where the first two factors can be further expanded as
$$
\Big(\bar\alpha\,\overline{\scalprod{\sigma'_1}{C_1 \,+}}+\bar\beta\,\overline{\scalprod{\sigma'_1}{C_1 \,-}}\Big)
\Big(\alpha {\scalprod{\sigma_1}{C_1 \,+}}+\beta{\scalprod{\sigma_1}{C_1 \,-}}\Big).
$$
By independence, the expectation factorizes and, with the notation $\scalprod{\sigma}{C \tau} =C_{\sigma,\tau}$, $\sigma, \tau\in\{+1, -1\}$, we immediately get from \fer{trc} and \fer{zcor} 
\begin{eqnarray*}
&&\E(\overline{C_{\sigma'_j,\sigma'_{j-1}}}C_{\sigma_j,\sigma_{j-1}})=0 \ \ \mbox{if }\ \ \sigma'_j\neq \sigma_{j}, \ \ \forall \ j\geq 2.
\end{eqnarray*}
This, together with the conditions $\sigma_n=\sigma'_{n}$ and $ \sum_{j=1}^n\sigma_j=\sum_{j=1}^n\sigma'_j=k$ in \fer{comp}, imposes $\sigma_j=\sigma'_{j}$ for all $j\geq 1$.
Hence, \fer{comp} reduces to
\bea
&&w_k(n)=\\ \nonumber
&&\sum_{\sigma \in \{-1,+1\}^n\atop \sum_{j=1}^n\sigma_j=k }
\E(|\alpha|^2|C_{\sigma_1,+1}|^2+|\beta|^2|C_{\sigma_1,-1}|^2+2\Re(\bar\alpha \beta
C_{\sigma_1,-1}\overline{C_{\sigma_1,+1}} ) )
\prod_{j=2}^n \E(|C_{\sigma_j,\sigma_{j-1}}|^2),
\eea
where 
\bea
&&\E(|C_{+1,+1}|^2)=\E(|C_{-1,-1}|^2)=t^2\nonumber\\
&&\E(|C_{+1,-1}|^2)=\E(|C_{-1,+1}|^2)=r^2\nonumber\\
&&\E(C_{-1,-1}\overline{C_{-1,+1}})=-\E(C_{+1,-1}\overline{C_{+1,+1}})=rt.
\eea
The result then follows with the definition of $S_n=\sum_{j=1}^n\sigma_j$.
\ep\\

Persistent random walks or correlated random walks are well known and have been studied in many details and greater generality,  ours being the simplest instance. See e.g. \cite{rh}, \cite{cr} and the references therein. In particular, when $r=t=1/\sqrt2$, the persistent and symmetric random walks are equivalent. Also it is known that 
the first moment is finite and that the second moment is proportional to $n$ for $n$ large. This leads to the following diffusive behaviour:
\bea
&&\lim_{n\ra\infty}\frac{\E(\bra X \ket_{\psi_0}(n))}{n}=0\nonumber\\
&&\lim_{n\ra\infty}\frac{\E(\bra X^2 \ket_{\psi_0}(n))}{n}=t^2/r^2.
\eea

For completeness, we provide below a simple proof of the fact that all moments of the persistent random walk display a diffusive behaviour, a statement that we couldn't find as such in the literature, although certainly well known. 

\medskip
{\bf Proof of Proposition \ref{diff}.}
We assume the hypotheses of Proposition \ref{prw} and use the familiar setup of generating functions together with a classical scaling argument.\\
We consider only the situation $r\neq t$ which differs from the usual symmetric random walk. Let $w_k^\pm(n)$ be the conditional probabilities
$$
w_k^\pm(n)=\P(S_n=k|\sigma_n=\pm1) \  \ \ \ \mbox{s.t.}\ \ \ \ w_k^+(n)+w_k^-(n)=w_k(n).
$$
Thus we have, for $n\geq 1$ and $|k|\leq n$
\bea\label{rec}
w_k^+(n+1)&=&r^2w_{k-1}^-(n)+t^2w_{k-1}^+(n)\nonumber\\
w_k^-(n+1)&=&t^2w_{k+1}^-(n)+r^2w_{k+1}^+(n)
\eea
with 
\be
w^+_1(1)=a, \ \ \ w^-_1(1)=b.
\ee
Moreover, $w_k^\pm(n)=0$ if $|k|>n$. We introduce the generating functions $\Phi^\pm_n$ and $\Psi_n$ by
\bea
\Phi^\pm_n(z)&=&\sum_{k=-n}^ne^{izk}w_k^\pm(n), \nonumber\\
\Psi_n(z)&=&\Phi^+_n(z)+\Phi^-_n(z), \ \ \forall z\in\C.
\eea
As a consequence of \fer{rec}, introducing $\Phi_n(z)=(\Phi^+_n(z),\Phi^-_n(z))^T$, we have
\bea
&&\Phi_{n+1}(z)=M(z)\Phi_n(z), \ \ \ \mbox{with}\\ \nonumber
&&M(z)= \left[\begin{array}{cc}
t^2e^{iz}&r^2e^{iz} \\ 
r^2e^{-iz}& t^2e^{-iz}
\end{array}
\right]\ \ \ \mbox{and}\ \ 
\Phi_{1}(z)=\left[ \begin{array}{c}
ae^{iz} \\ 
be^{-iz}
\end{array}
\right].
\eea
This allows to determine explicitly $\Phi_n(z)$ and 
\be
\Psi(z)=\scalprod{\left[\begin{array}{c} 1\\ 1\end{array}\right]}{M^{n-1}(z)\Phi_1(z)}, \ \ \ n\geq 1,
\ee
and, in turn, all moments of the probability distribution $\{w_k(n)\}_{k\in\Z}$. \\

We consider now the diffusive scaling introducing  the macroscopic time variable $N=n\tau$,  where $\tau >>1$,  $\tau\in\N$ and the macroscopic space variable 
$K=\sqrt\tau k$, such that $K/\sqrt{N}=k/\sqrt{n}$ remains finite. Expecting a probability distribution $\{w_k(n)\}_{k\in\Z}$ asymptotically invariant under this scaling, we are led to the study of the generating function at $z=y/\sqrt\tau$ in the limit $\tau\ra\infty$:
\begin{lem}
\label{l4.4}
\be
\lim_{\tau\ra\infty\atop\tau\in\N}\sum_{k=-\tau n}^{\tau n}e^{i\frac{y}{\sqrt\tau}k}w_k(\tau n)=e^{-n\frac{t^2}{2r^2}y^2},
\label{mmmmm}
\ee
uniformly in $y$ in any compact set of $\C$.
\end{lem}
Since the functions of $y$ involved are entire and the convergence is uniform in compact sets, we can differentiate the above identity w.r.t. $y$ as many times as we wish. In particular, differentiating $L$ times with $n=1$ and setting $y=0$ immediately yields the following Corollary which ends the proof of the Proposition \ref{diff}:
\begin{cor} 
\be
\lim_{\tau\ra\infty\atop\tau\in\N}\sum_{k=-\tau }^{\tau }\frac{(ik)^L}{\tau^{L/2}} w_k(\tau)=\left(\frac{t^2}{r^2}\right)^{L/2}H_L(0)(-1)^L,
\ee
where $H_L$ denotes the Hermite polynomial $H_L(z)=(-1)^Le^{z^2/2}\left(\frac{d}{dz}\right)^Le^{-z^2/2}$.
\end{cor}
{\bf Proof of Lemma \ref{l4.4}.\ }
Note that the left side of \fer{mmmmm} equals
\be
\lim_{\tau\ra\infty}\Psi_{\tau n}(y/\sqrt{\tau})=\lim_{\tau\ra\infty}\scalprod{\left[\begin{array}{c} 1\\ 1\end{array}\right]}{M^{\tau n-1}(y/\sqrt{\tau})\Phi_1(y/\sqrt{\tau})}
\ee
so that we are interested in small values of $z=y/\sqrt{\tau}$.
For $|z|$ small enough, we compute the spectrum of the analytic matrix $M(z)$
\be
\sigma(M(z))=\{\lambda_1(z), \lambda_2(z)\}
\ee
with distinct eigenvalues
\be
\lambda_{1\atop 2}(z)=t^2\cos(z)\pm\sqrt{t^4\cos^2(z)+1-2t^2}.
\ee
Hence, for $|z|$ small,  there exists an invertible matrix $R(z)$, analytic in $z$, such that
\be
M^n(z)=R^{-1}(z)\left[\begin{array}{cc}\lambda^n_{1}(z)& 0 \\0& \lambda^n_{2}(z)\end{array}\right]R(z).
\ee
For $|z|$ small enough, we have
\bea\label{evz}
\lambda_{1}(z)&=&1-z^2\frac{t^2}{2r^2}+O(z^4)\nonumber \\ 
\lambda_{2}(z)&=&(t^2-r^2)+z^2\frac{t^2(t^2-r^2)}{2r^2}+O(z^4),
\eea
whereas
$$M^{-1}(0)=M^{-1}(0)^*=\frac1{t^2-r^2}\left[\begin{array}{cc}
t^2&-r^2\\-r^2&t^2\end{array}\right] \ \ \ \mbox{satisfies}\ \  \
M^{-1}(0)\left[\begin{array}{cc} 1\\ 1\end{array}
\right]=\left[\begin{array}{c}1\\ 1\end{array}\right],$$
and 
$$
R(0)=R^*(0)=R^{-1}(0)=\frac{1}{\sqrt2}\left[\begin{array}{cc}1&1\\ 1&-1\end{array}\right].
$$
Since \fer{evz} yields for any $y\in\C$ and $\tau $ large enough
\bea
\lambda_{1}(y/\sqrt{\tau})^{\tau n}&=&\e^{-ny^2\frac{t^2}{2r^2}}+O(ny^4/\tau)\nonumber\\
\lambda_{2}(y/\sqrt{\tau})^{\tau n}&=&(t^2-r^2)^{\tau n}\e^{ny^2\frac{t^2}{2r^2}}+O(ny^4/\tau),
\eea
we eventually obtain, since $|r^2-t^2|<1$ and $a+b=1$,
\bea
&&\lim_{\tau\ra \infty}\scalprod{\left[\begin{array}{c} 1\\ 1\end{array}\right]}{ M^{\tau n-1}(y/\sqrt{\tau})\Phi_1(y/\sqrt{\tau})}=
\\ \nonumber
&&\qquad\qquad\qquad\qquad
\scalprod{\left[\begin{array}{cc}e^{-ny^2\frac{t^2}{2r^2}}& 0 \\ 0 &0\end{array}\right] \left[\begin{array}{c}1\\ 0\end{array}\right]}{ \left[\begin{array}{c}a+b\\ a-b\end{array}\right] }=\e^{-ny^2\frac{t^2}{2r^2}}
\eea
\ep

\appendix


\renewcommand{\theresultcounter}{\Alph{section}.\arabic{resultcounter}}


\section{Appendix}

\subsection{Proof of Lemma \ref{ballistic}}

Introducing the discrete Fourier transform $\cF: L^2([0,2\pi),\C^2)\ra \C^2\otimes l^2(\Z)$ by
\be
(\cF f)(k)=\hat f(k)=\frac{1}{2\pi}\int_0^{2\pi}f(x)e^{-ikx}dx,
\ee
we get that $U$ is unitarily equivalent to a multiplication operator by a matrix $V(x)$.
\be\label{defv}
\cF^{-1}U\cF=\mbox{mult}V(x)=\mbox{mult}\left[\begin{array}{cc}e^{-ix}& 0\\ 0&e^{ix}\end{array}\right]C.
\ee
The matrix $V(x)$ is entire in $x$, and unitary for $x$ real, which means that its eigenvalues $\{\alpha_1(x),\alpha_2(x)\}$ and eigenprojectors $\{P_1(x), P_2(x)\}$ are also analytic functions of $x\in\R$, even at the possible crossings of eigenvalues for $x\in\R$. Moreover,   for any $n\in\Z$, 
\be
V^n(x)=P_1(x)\alpha_1^n(x)+P_2(x)\alpha_2^n(x).
\ee
Similarly, the position operator $K=(\I\otimes k)$ is unitarily equivalent to differentiation w.r.t. $x$ (on its natural domain):
\be
\cF^{-1}K\cF=-i\partial_x.
\ee
In particular, for $\Psi\in\C^2\otimes l^2(\Z)$ such that $\cF^{-1}\Psi=f\in L^2([0,2\pi),\C^2)$ and $\Psi$ in the domain of $K^2=(\I\otimes k)^2$, we have for all $n\in\Z$,
\be
\bra K^2\ket_{\Psi}(n):=\scalprod{\Psi}{U^{-n}K^2U^n\Psi}=\int_{0}^{2\pi}\|\partial_x (V^n(x)f(x))\|^2_{\C^2}dx.
\ee
Now, it is not difficult to see by explicit computations that this quantity behaves as $n^2$ for $n\ra \infty$, unless the analytic eigenvalues $\{\alpha_1(x),\alpha_2(x)\}$ of $V(x)$ are independent of $x$. We have 
\be
\alpha_1(x)\alpha_2(x)=\det V(x)=\det C \ \ \ \mbox{and}\ \ \ \alpha_1(x)+\alpha_2(x)=\mbox{tr } V(x)=e^{-ix}a+e^{ix}d.
\ee
Hence the eigenvalues $\alpha_j(x)$ are independent of $x$ iff $a=d=0$. \ep

\subsection{Positivity and continuity of Lyapunov exponent}

This Section is devoted to the proof of Theorem \ref{thmlyap}.\\

Remember that we take $\mu$ to be absolutely continuous with density $\tau\in L^\infty(\T)$ having support with nonempty interior and that the transfer matrices $T_z(\theta,\eta)$ are given in \fer{m9} (with $\omega_{2n}=\theta$ and $\omega_{2n-1}=\eta$).

To prove Theorem \ref{thmlyap}, we use Furstenberg's theorem, which is developed for real square matrices. We thus map $T_z\in {\mathbb M}_2(\cx)$ (square $2\times 2$ matrices with complex entries) into $\tau(T_z)\in{\mathbb M}_4(\rx)$ using the bijection
\begin{equation}
\tau: 
\left[
\begin{array}{cc}
a & b \\
c & d
\end{array}
\right]
\mapsto 
\left[
\begin{array}{cc}
I {\rm Re} a  +J{\rm Im}a  & I {\rm Re} b + J{\rm Im}b \\
I {\rm Re} c  +J{\rm Im}c  & I {\rm Re} d + J{\rm Im}d
\end{array}
\right],
\end{equation}
where 
$$
I = \left[
\begin{array}{cc}
1 & 0 \\
0 & 1
\end{array}
\right] \mbox{\ and\ }
J = \left[
\begin{array}{cc}
0 & 1 \\
-1 & 0
\end{array}
\right].
$$
We refer to \cite{BHJ} for more detail on this transformation. In particular, we have  
\begin{equation}
\|\tau(A)\| = \sqrt{2}\|A\|,
\label{m66}
\end{equation}
where the norms are given by $\|X\|^2 ={\rm Tr}(X^*X)$, for $X\in{\mathbb M}_2(\cx)$ or $X\in{\mathbb M}_4(\rx)$, and that if $A\in{\mathbb M}_2(\cx)$ satisfies $|\det(A)|=1$, then $|\det\tau(A)|=1$. Due to \fer{m67}, we have $|\det\tau(T_z(\theta,\eta))|=1$, for all $z\neq 0$ and all $\theta,\eta$. 

Relation \fer{m66} together with the fact that $\tau(AB)=\tau(A)\tau(B)$ shows that the statement of Theorem \ref{thmlyap} is equivalent to $\lim_{n\rightarrow\infty}\frac{1}{n}\|\tau(T_z(\theta_n,\eta_n))\cdots \tau(T_z(\theta_1,\eta_1))\| =\gamma$ almost surely, for the same deterministic $\gamma>0$ as in Theorem \ref{thmlyap}.

\medskip

The measure $\mu$ on ${\mathbb T}$ induces a measure on ${\mathbb M}_4({\mathbb R})$, supported on the subset 
\begin{equation}
{\cal M}:=\{\tau(T_z(\theta,\eta)), \theta,\eta\in {\rm supp}\mu\}.
\label{m71}
\end{equation}
We call the induced measure again $\mu$. Let $Y_1,Y_2,\ldots$ be iid random matrices in ${\mathbb M}_4(\rx)$ with common distribution $\mu$. If the integrability condition 
\begin{equation}
\label{finitelog}
\E\left[\max\{\log\|Y_1\|, 0\}\right]<\infty
\end{equation}
is satisfied, then the upper Lyapunov exponent $\gamma\in\rx\cup\{-\infty\}$ is defined as 
$$
\gamma:=\lim_{n\rightarrow\infty} \frac{1}{n}\E[\log\|Y_n\cdots Y_1\|].
$$
The theorem of Furstenberg and Kesten (\cite{BL}, Theorem 4.1) states that if in addition the matrices $Y_j$ are invertible ($Y_j\in{\rm GL}_4(\rx)$), then
\begin{equation}
\lim_{n\rightarrow\infty}\frac{1}{n}\log\|Y_n\cdots Y_1\|=\gamma
\label{m70}
\end{equation}
almost surely. In our case, $Y_j=\tau(T_z(\theta_j,\eta_j))$ is invertible, and $\|Y_1\|\leq C$ is uniformly bounded in $\theta_1,\eta_1$ (see \fer{m9}), so that \fer{finitelog} is trivially satisfied. This implies that \fer{m69} holds almost surely, with a deterministic $\gamma$. Moreover, since $S=T_z(\theta_n,\eta_n)\cdots T_z(\theta_1,\eta_1)$ is a $2\times 2$ invertible matrix, we have $\|S\|=\|S^{-1}\|\geq 1$, and therefore $\gamma\geq 0$. The remaining part of the proof of Theorem \ref{thmlyap} consists in proving that $\gamma$ is {\it strictly} positive and continuous.

\medskip

Let ${\cal G}_\mu \subset {\rm GL}_4(\rx)$ be the (multiplicative) group of matrices generated by the transfer matrices $T_z(\theta,\eta)$, where $\theta,\eta$ vary throughout the support of the measure $\mu$. Here, $z$ is fixed and not displayed in ${\cal G}_\mu$.

\begin{thm}[Furstenberg, \cite{BL} Thm. 6.3]
\label{thmfurstenberg}
Suppose that ${\cal G}_\mu$ is strongly irreducible and non-compact. Then the upper Lyapunov exponent associated with any sequence of random matrices $Y_1, Y_2,\ldots$ in ${\rm SL}_4(\rx)\cap {\rm supp}\mu$,  iid  with common distribution $\mu$, is strictly positive. This means that \fer{m70} holds with $\gamma>0$. 
\end{thm}

Now we show that ${\cal G}_\mu$ is strongly irreducible and non-compact.

\begin{lem}
\label{lemma3}
If ${\rm supp}\mu$ contains two distinct points, then ${\cal G}_\mu$ is not compact for all $z\neq 0$.
\end{lem}

{\it Proof.\ } For fixed $z=R\e^{\i\alpha}$, we have 
$$
T_z(\theta,\eta) = 
\frac{\e^{-\i(\theta+\alpha)}}{R} 
\left[
\begin{array}{cc}
R^2\e^{\i(\theta+\alpha+\eta+\alpha)} +r^2 & -rt\\
-rt & t^2
\end{array}
\right],
$$
and a direct calculation shows that 
$$
T_z(\eta,\theta) T_z(\eta,\eta)^{-1}=(T_z(\theta,\theta)^{-1} T_z(\theta,\eta))^* =
\left[
\begin{array}{cc}
\e^{\i(\theta-\eta)} & \textstyle\frac rt(\e^{\i(\theta-\eta)} -1)\\
0 & 1
\end{array}
\right].
$$
Both operators $T_z(\eta,\theta) T_z(\eta,\eta)^{-1}$ and $T_z(\theta,\theta)^{-1} T_z(\theta,\eta)$ belong to ${\cal G}_\mu$, and hence so does their positive-definite product,
$$
0\leq M:=T_z(\eta,\theta) T_z(\eta,\eta)^{-1} T_z(\theta,\theta)^{-1} T_z(\theta,\eta) = \bbbone +\frac rt
\left[
\begin{array}{cc}
\textstyle\frac rt |\e^{\i(\theta-\eta)}-1| & \e^{\i(\theta-\eta)} -1\\
\e^{-\i(\theta-\eta)}-1 & 0
\end{array}
\right].
$$
Note that the r.h.s. does not depend on $z$ at all. Since the determinant of the last matrix is $-\frac{r^2}{t^2}|\e^{\i(\theta-\eta)}-1|^2<0$, ${\cal G}_\mu$ contains a non-negative matrix $M$ having an eigenvalue strictly larger than one. It follows that $\|M^n\|\rightarrow\infty$ as $n\rightarrow \infty$. Thus the group generated by the $\tau(T_z(\theta,\eta))$, $\theta,\eta\in{\rm supp}\mu$, is not compact. \hfill $\blacksquare$

\medskip

Finally we prove strong irreducibility of ${\cal G}_\mu$. Let $J$ be a non-empty open interval, $J\subset{\rm supp}\mu$. Since the subset of matrices ${\cal M}'$ obtained from $\cal M$, \fer{m71}, by restricting $\theta,\eta\in J$, is a continuous image of the connected set $J\times J\subset \rx^2$, ${\cal M}'$ is a connected set in ${\mathbb M}_4(\rx)$. Strong irreducibility of ${\cal M}'$ (which implies strong irreducibility of $\cal M$) is then equivalent to irreducibility of ${\cal M}'$, see \cite{BL} Exercise IV.2.9.

\begin{lem}
\label{lemirred}
The only subspaces $V\subseteq\rx^4$ invariant under the action of ${\cal M}'$ are $V=\{0\}$ and $V=\rx^4$. Hence ${\cal M}'$ is strongly irreducible. Since ${\cal M}'\subset {\cal G}_\mu$, we have that ${\cal G}_\mu$ is strongly irreducible. 
\end{lem}

{\it Proof.\ }
We just need to show that $V=\{0\}$ and $V=\rx^4$ are the only invariant subspaces of $\cal M'$. Let $z=R\e^{\i\alpha}\neq 0$ be fixed. One easily finds that 
\begin{equation}
\tau(T_z) = \frac{1}{R} \left[ T_1\cos\varphi +T_2\sin\varphi +T_3\cos\chi +T_4\sin\chi \right],
\end{equation}
where $\varphi=\alpha+\eta$, $\chi=\alpha+\theta$ vary in the open interval $\alpha+J$, and the matrices $T_j$ are given by
\begin{equation*}
T_1=R^2 \left[
\begin{array}{cccc}
1 & 0 & 0 & 0 \\
0 & 1 & 0 & 0\\
0 & 0 & 0 & 0\\
0 & 0 & 0 & 0
\end{array}
\right], 
\qquad 
T_2 = R^2
 \left[
\begin{array}{cccc}
0 & 1 & 0 & 0 \\
-1 & 0 & 0 & 0\\
0 & 0 & 0 & 0\\
0 & 0 & 0 & 0
\end{array}
\right],
\end{equation*}
\begin{equation*}
T_3= \left[
\begin{array}{cccc}
r^2 & 0 & -rt & 0 \\
0 & r^2 & 0 & -rt\\
-rt & 0 & t^2 & 0\\
0 & -rt & 0 & t^2
\end{array}
\right], 
\qquad 
T_4 =
 \left[
\begin{array}{cccc}
0 & -r^2 & 0 & rt \\
r^2 & 0 & -rt & 0\\
0 & rt & 0 & -t^2\\
-rt & 0 & t^2 & 0
\end{array}
\right].
\end{equation*}

By taking derivatives in the angles, we see that if $V$ is invariant under $\cal M'$, then $V$ is invariant also under the action of $-T_1\sin\varphi+T_2\cos\varphi$ (and $-T_3\sin\chi+T_4\cos\chi$), and hence (by again differentiating) $V$ is as well invariant under the action of $T_1\cos\varphi +T_2\sin\varphi$ (and $T_3\cos\chi+T_4\sin\chi$). Therefore, $-T_1\sin^2\varphi +T_2\sin\varphi\cos\varphi$ and $T_1\cos^2\varphi+T_2\sin\varphi\cos\varphi$ leave $V$ invariant, and hence so does $T_1$ (the difference) and consequently $T_2$ as well. Similarly one sees that $T_3, T_4$ leave $V$ invariant, too. Hence any subspace $V$ invariant under $\cal M'$ must be invariant separately under $T_1$, $T_2$, $T_3$ and $T_4$.

Consider first $V$ with $\dim V=1$, i.e., $V=\langle v\rangle$ (real span). It is easy to see that the only possibility for $v$ resulting in a $V$ invariant under $T_1+T_2$ and $T_1-T_2$ is $v=\alpha e_3+\beta e_4$ (canonical basis elements of $\rx^4$). But either such $V$ is not left invariant under the action of $T_3$. Consequently, no one-dimensional subspace is invariant under $\cal M$. 

Consider next $V$ with $\dim V=2$. Since $T_3$ is real symmetric and must leave $V$ invariant, $V$ must be spanned by two eigenvectors of $T_3$. One easily finds that $T_3$ has eigenvalues $\{0,1\}$, both twice degenerate, that a basis for the kernel of $T_3$ is $\{ [1, 0, r/t, 0]^t, [0,1,0,r/t]^t\}$ (transpose), and a basis of the eigenspace with eigenvalue 1 is $\{[1,0,-t/r,0]^t, [0,1,0,-t/r]^t\}$. There are three possible cases: 1. both eigenvectors spanning $V$ belong to the kernel of $T_3$, 2. both  eigenvectors spanning $V$ belong to the eigenspace with eigenvalue 1 of $T_3$, or 3. one eigenvector belongs to the kernel of $T_3$ and the other one belongs to the other spectral subspace. Either of these three cases can be analyzed separately, and one finds that none of the thus formed spaces $V$ with $\dim V=2$ is invariant under all of the $T_j$, $j=1,2,3,4$. In conclusion, no two-dimensional subspace is invariant under $\cal M$. 

Consider now $V$ with $\dim V=3$. Then $V^\perp$ has dimension 1 and is invariant under $T_3$, since the latter is real symmetric. Hence $V^\perp$ is spanned by one of the eigenvectors of $T_3$. In the same way, one sees that $V^\perp$ must also be invariant under $T_1$, and one finds easily that this implies that $V^\perp=\{0\}$, a contradiction to $\dim V=3$. This shows that there is no three-dimensional subspace invariant under $\cal M$. Lemma \ref{lemirred} follows. \hfill $\blacksquare$

\medskip
This proves all assertions of Theorem \ref{thmlyap}, except for the continuity of $z\mapsto \gamma(z)$. However, the latter has been shown to hold in Section VII of \cite{HS} (Theorem 7.1).\hfill $\blacksquare$

\end{document}